%% file: degen_neutrinos.tex
\documentclass[a4paper,12pt]{article}

\input{pack-def}
\usepackage[small,sc]{caption}
\usepackage{booktabs}

\begin{document}
\onehalfspacing

\begin{titlepage}

\vspace*{-15mm}
\begin{flushright}
TTP14-034
\end{flushright}
\vspace*{0.7cm}

\begin{center} {
    \bfseries\LARGE
    Radiative generation of neutrino mixing:
    degenerate masses and threshold corrections}
\\[8mm]
Wolfgang~Gregor~Hollik
\footnote{E-mail: \texttt{wolfgang.hollik@kit.edu}}
\\[1mm]
\end{center}
\vspace*{0.50cm}
\centerline{\itshape
Institut f\"ur Theoretische Teilchenphysik, Karlsruhe Institute of Technology}
\centerline{\itshape
Engesserstra\ss{}e 7, D-76131 Karlsruhe, Germany }
\vspace*{1.20cm}

\begin{abstract}
\noindent
Degenerate neutrino masses are excluded by experiment. The
experimentally measured mass squared differences together with the yet
undetermined absolute neutrino mass scale allow for a quasi-degenerate
mass spectrum. For the lightest neutrino mass larger than roughly
\(0.1\,\eV\), we analyze the influence of threshold corrections at the
electroweak scale. We show that typical one-loop corrections can
generate the observed neutrino mixing as well as the mass differences
starting from exactly degenerate masses at the tree-level. Those
threshold corrections have to be explicitly flavor violating. Flavor
diagonal, non-universal corrections are not sufficient to simultaneously
generate the correct mixing and the mass differences. We apply the new
insights to an extension of the Minimal Supersymmetric Standard Model
with non-minimal flavor violation in the soft breaking terms and
discuss the low-energy threshold corrections to the light neutrino mass
matrix in that model.
\end{abstract}
{\footnotesize PACS: 14.60.Pq, 12.60.Jv}
\end{titlepage}

\setcounter{footnote}{0}

\section{Introduction}
A direct measurement of neutrino masses is still missing. The
perspective of data from tritium decay in the near future probes the
effective electron neutrino mass down to
\(0.2\,\eV\)~\cite{Osipowicz:2001sq}. Complementary to direct searches
are constraints from cosmology~\cite{Abazajian:2011dt,Abazajian:2013oma}
where the tightest bound on the sum of (active) neutrino masses released
by the Planck collaboration is under certain assumptions \(\sum m_\nu <
0.23\,\eV\)~\cite{Ade:2013zuv}. Including galaxy clustering and lensing
data in addition to the standard observation from the cosmic microwave
background and baryon acoustic oscillations, this bound turns into an
observation in a ``degenerate active neutrino scenario'', \(\sum m_\nu =
(0.320\pm0.081)\,\eV\)~\cite{Battye:2013xqa}. The reported sum of
neutrino masses shows a significant deviation of about \(3.2\,\sigma\)
from the minimal value \(\sum m_\nu = 0.059\,\eV\) which is allowed by a
massless lightest neutrino and the differences of mass squares:
\begin{equation}\label{eq:Deltam2}
\begin{aligned}
\Delta m_{21}^2 &= 7.50^{+0.19}_{-0.17} \times 10^{-5}\,\eV^2, \\
\Delta m_{31}^2 &= 2.457\pm 0.047 \times 10^{-3}\,\eV^2,
\end{aligned}
\end{equation}
where \(\Delta m_{ji}^2 = m_j^2 - m_i^2\) and we restricted ourselves
to the result of a normal hierarchy (\(\Delta m_{31}^2 > 0\)) as follows
from a global fit of neutrino oscillation data~\cite{Gonzalez-Garcia:2014bfa}.

The cosmological bounds as stated above disfavor a possible direct
detection of a neutrino mass from tritium decay. In any case, presuming
such a discovery (which would be around \(m_{\nu_e} = 0.35\,\eV\) or
higher) or taking the cosmological observation \(\sum m_\nu =
0.32\,\eV\) for granted, we observe a neutrino mass spectrum that is
(quasi-)degenerate. In the first case, the three neutrino masses differ
only about one percent. The second scenario has at least the third mass
about ten percent larger than the lightest and second-lightest.

The three neutrino mass eigenvalues are calculated, once the absolute
scale \(m_0\) is fixed (we focus on a normal mass hierarchy):
\[
\left| m_1 \right| = m_0 \,,\qquad\qquad
\left| m_2 \right| = \sqrt{m_0^2 + \Delta m_{21}^2} \,,\qquad\qquad
\left| m_3 \right| = \sqrt{m_0^2 + \Delta m_{31}^2}.
\]
Now we see, that \(m_{1\ldots 3}\) are basically the same numbers, in
the limit \(m_0^2 \gg \Delta m^2\). The striking feature of such a
quasi-degenerate mass spectrum is, that the small deviations from exact
degeneracy can be seen as a small perturbation originating in quantum
corrections to neutrino masses.

It is well-known, that for degenerate neutrinos quantum corrections are
important~\cite{Ellis:1999my, Casas:1999tp, Casas:1999ac, Haba:1999xz,
  Chun:1999vb, Balaji:2000gd, Chankowski:2000fp, Chun:2001kh,
  Chankowski:2001mx, Mohapatra:2003tw, Brahmachari:2003dj,
  Mohapatra:2005gs, Haba:2012ar}. However, the question is, whether the
stringent Planck bounds still allow for dominant contributions from
quantum corrections. We shall focus on the influence of low-energy
threshold corrections as they may arise in several extensions of the
Standard Model around the \(\TeV\) scale. The importance of threshold
corrections in the MSSM was already pointed out~\cite{Chun:1999vb,
  Chankowski:2000fp, Chun:2001kh, Chankowski:2001mx, Mohapatra:2005gs},
we apply them to the current phenomenology of neutrino masses and mixing
(simultaneously accommodating a non-zero third mixing angle and still
being consistent with the mass bound).

Furthermore, those loop corrections to neutrino masses may also lead to
the desired mixing pattern as will be shown. The observed values for the
three mixing angles are taken from the same global fit that determines
the \(\Delta m^2\) \cite{Gonzalez-Garcia:2014bfa}
\begin{equation}\label{eq:measured-angles}
\begin{aligned}
\sin^2\theta_{12} &= 0.304 \pm 0.012, \\
\sin^2\theta_{13} &= 0.0219^{+0.0010}_{-0.0011}, \\
\sin^2\theta_{23} &= 0.451 \pm 0.001,
\end{aligned}
\end{equation}
where we follow the standard parametrization of any three-dimensional
rotation matrix that can be seen as three successive rotations and the
Dirac \(\CP\) phase \(\delta_\CP\) associated to the 1-3-rotation
\begin{equation}\label{eq:standardparam}
\begin{aligned}
  \Mat{U}(\theta_{12},\theta_{13},\theta_{23},\delta_\CP,\alpha_1,\alpha_2)
  \;&=\;
  \Mat{U}_{23}(\theta_{23}) \; \Mat{U}_{13}(\theta_{13}, \delta_\CP)
  \; \Mat{U}_{12}(\theta_{12}) \; \Mat{P}(\alpha_1,\alpha_2) \\
  &= \begin{pmatrix}
    c_{12} c_{13} & s_{12} c_{13} & s_{13} e^{-\im\delta_\CP} \\
    - s_{12} c_{23} - c_{12} s_{23} s_{13} e^{\im\delta_\CP} & c_{12} c_{23}
    - s_{12} s_{23} e^{\im\delta_\CP}
    s_{13} & s_{23} c_{13} \\
    s_{12} s_{23} - c_{12} c_{23} s_{13} e^{\im\delta_\CP} & - c_{12} s_{23}
    - s_{12} c_{23} s_{13} e^{\im\delta_\CP} & c_{23} c_{13}
\end{pmatrix} \; \Mat{P},
\end{aligned}
\end{equation}
where \(s_{ij} = \sin\theta_{ij}\) and \(c_{ij} = \cos\theta_{ij}\) and
the angles \(\theta_{ij}\) parametrize the rotation in the \(i\)-\(j\)
plane. The phase matrix \(\Mat{P} =
\diag(e^{\im\alpha_1},e^{\im\alpha_2}, 1)\) contains the Majorana phases
which are absent in the case of Dirac neutrinos.

We work in a basis, where the charged lepton Yukawa couplings are
diagonal and generate the small neutrino masses via a seesaw-inspired
model \cite{Minkowski:1977sc,Mohapatra:1979ia, Yanagida:1980xy,
  GellMann:1980vs, Schechter:1980gr, Magg:1980ut, Schechter:1981cv} that
leads to a nonrenormalizable operator suppressed by a reasonably heavy
scale \(M\) \cite{Weinberg:1979sa}:
\begin{equation}\label{eq:WeinbergOp}
  \mathcal{L}^\ell_\upsh{Y} \supset Y^\el_{\alpha} L_\alpha \cdot H\; e_\alpha
  + \frac{h^{\alpha\beta}}{M} (L_\alpha \cdot H) (L_\beta \cdot H)
  + \;\hc,
\end{equation}
where \(H\) denotes the SM Higgs doublet whose neutral component
acquires a \emph{vev} and \(L_\alpha\) is the lepton doublet of flavor
\(\alpha = \el,\mul,\taul\): \(L_\alpha = (\nu_{\Ll,\alpha},
\ell_{\Ll,\alpha})\) of left-handed fields and \(e_\alpha\) denotes the
right-handed charged leptons. If the couplings \(h^{\alpha\beta}\) are
\(\mathcal{O}(1)\) couplings, the mass of the light neutrinos scales as
\(v^2/M\). To end up with \(m_\nul \lesssim 1\,\eV\), the scale \(M\)
has to be \(\mathcal{O}(10^{14}\,\GeV)\).

We therefore explicitly work with Majorana neutrinos, which is manifest
in a symmetric neutrino mass matrix, \(h^{\alpha\beta} =
h^{\beta\alpha}\), and the Majorana phase matrix \(\Mat{P}\) of
Eq.~\eqref{eq:standardparam}. Since the charged lepton masses are
diagonal, \(\Mat{U}\) of Eq.~\eqref{eq:standardparam} diagonalizes the
neutrino mass matrix and shows up as mixing matrix in the weak charged
current interaction which is known as Pontecorvo-Maki-Nakagawa-Sakata
(PMNS) matrix~\cite{Pontecorvo:1957qd, Maki:1962mu}.

A Majorana mass term does not allow to rotate away complex phases into
redefinitions of the fields. Majorana neutrino masses are in general
complex and the two Majorana phases \(\alpha_{1}, \alpha_{2}\) can be
absorbed in a redefinition of the masses:
\begin{equation}\label{eq:diagMajMass}
\bar{\Mat{m}}^\nu = \Mat{U}^\tp \Mat{m}^\nu \Mat{U}
 = \diag( |m_1| e^{-2\im\alpha_1}, |m_2| e^{-2\im\alpha_2}, m_3 ),
\end{equation}
where \(\Mat{U}\) is the mixing matrix of Eq.~\eqref{eq:standardparam}
without the phase matrix \(\Mat{P}\). Without loss of generality
\(m_3\) can be chosen real and positive.  Under \(\CP\) conservation,
the choice \(\alpha_{1,2} = 0, \pm \frac{\pi}{2}\) results in the
possibility of having two different signs for the masses \(m_{1,2}\)
reflecting different \(\CP\) parities. In the following, we want to
discuss the two scenarios where all three neutrinos have the same
\(\CP\) parity and where one has a different sign.

This paper is organized as follows: in the second section we discuss
possible patterns of degenerate masses as was done in the literature
some time ago. In earlier works, people focused on bi-maximal (or
tri-bimaximal) tree-level mixing and considered deviations from that
pattern, especially with only very small \(\theta_{13}\). Nowadays,
since \(\theta_{13}\) is measured quite well, it is intriguing to
re-examine exemplarily simple patterns in the threshold corrections and
see whether they can deal with both rather large 1-3 mixing and the
established mass square differences. In the same section we also
consider the case with trivial mixing at tree-level and generate the
mixing fully radiatively. We estimate the size of the threshold
corrections needed to generate the observed mixing pattern. The same
procedure also fits the masses. Furthermore, in Section~\ref{sec:nuMSSM}
we give an explicit example of a supersymmetric model with exactly
degenerate tree-level masses, in which threshold corrections generate
the observed pattern of mass differences and mixings.

\section{The influence of Quantum corrections}
\label{sec:general}
In general, quantum corrections to neutrino masses change the alignment
pattern of mass eigenstates and mixing angles. A tree-level mass matrix
\(\Mat{m}^{(0)}_\nu\) provides a special mixing pattern by misalignment
of interaction (flavor) and mass eigenstates. This can be result of
some flavor symmetry acting at the tree-level Lagrangian. Quantum
corrections are, however, mixing different flavor eigenstates
differently from the tree-level pattern, where the corrected mass matrix
can be parametrized as
\begin{equation}\label{eq:loop_flav}
m^\nul_{\alpha\beta} = m^{(0)}_{\alpha\beta} +
m^{(0)}_{\alpha\gamma} I_{\gamma\beta} + I_{\alpha\gamma} m^{(0)}_{\gamma\beta},
\end{equation}
and \(I_{\alpha\beta}\) denote the corrections~\cite{Chun:1999vb,
  Chankowski:2000fp, Chun:2001kh, Chankowski:2001mx}. Greek indices
(\(\alpha,\beta,\gamma=\el,\mul,\taul\)) live in the interaction
basis. Note, that the tree-level mass matrix \(\Mat{m}^{(0)}\) as well
as the corrections \(\Mat{I}\) are symmetric in the case of Majorana
neutrinos which is explicitly assumed by the use of the effective
operator of Eq.~\eqref{eq:WeinbergOp}. The physical mass matrix is then
understood to be \(\Mat{m}^\nul\). In general, quantum corrections are
to be decomposed in contributions from the renormalization group (RG)
and low-energy threshold corrections at the electroweak scale: \(\Mat{I}
= \Mat{I}^\upsh{RG} + \Mat{I}^\upsh{TH}\).

The contributions from the renormalization group are known to give a
sizable effect for quasi-degenerate neutrino
masses~\cite{Chankowski:1993tx, Haba:1998fb, Ellis:1999my, Casas:1999tp,
  Casas:1999ac, Casas:1999tg, Chankowski:1999xc}. Especially the choice
of the same \(\CP\) parity for two masses may lead to large mixing at
low scale irrespective of the original mixing at the high scale
\cite{Balaji:2000gd}, known as infrared fixed
points~\cite{Chankowski:1999xc}. The effect from the renormalization
group severely depends on the Majorana phases: for a vanishing Majorana
phase, maximal mixing patterns get diluted on the way to the high scale
\cite{Haba:1999xz} for quasi-degenerate (\(m_0 \sim
\mathcal{O}(1\,\eV)\)) neutrino masses. If the phase on the contrary is
large or the overall mass scale much smaller than \(1\,\eV\), maximal
mixing is preserved. Likewise, zero mixing (as follows from the
assignment \(m_1 = m_2 = m_3\)) is conserved \cite{Haba:2000tx}
irrespective of the Majorana phase difference \(|\alpha_1 -
\alpha_2|\). We therefore safely neglect those kinds of contributions
that preserve specific mixing patterns anyway.

The main contribution to the mass splitting from the RG evolution is
from the \(\taul\) Yukawa coupling and can be estimated to be
\(I^\text{RG}_\taul \lesssim \frac{y_\taul^2}{16\pi^2}
\log\left(\frac{M}{M_Z}\right)\), where \(M\) is the heavy scale of
Eq.~\eqref{eq:WeinbergOp}. In the cosmological allowed scenarios (\(m_0
\lesssim 0.1\,\eV\)), the maximal splitting for large values of
\(\tan\beta\) in the MSSM and a heavy scale \(M \leq 10^{14}\;\GeV\) is
generically too small to generate the required splitting between \(m_3\)
and \(m_{1,2}\) which has to be around \(0.01\,\eV\), whereas
\(I_\taul^\text{RG} \approx 4.6\times 10^{-3}\) with \(\tan\beta=50\)
and \(M=10^{14}\,\GeV\). This correction, which is independent from the
tree-level neutrino mass spectrum, still gives a sizable effect for
\(m_0=0\,\eV\) of about \(10\%\) of the mass splitting \(\sqrt{\Delta
  m_{31}^2}\). To keep \(I_\taul^\text{RG}\) even smaller, one either
has to reduce \(M\) or \(\tan\beta\) or both. It is, however,
interesting to verify whether low-energy threshold corrections
themselves can account for the required mass splittings. Since for the
degenerate spectrum \(m_1^{(0)} = m_2^{(0)} = m_3^{(0)}\) zero mixing is
preserved, the existence of a RG induced mass splitting due to the
\(\tau\) Yukawa coupling does not alter the qualitative discussion on
that scenario below. This feature may be used to generate a larger
\(\Delta m_{31}^2\) compared to \(\Delta m_{21}^2\) if the the threefold
degeneracy is to be abandoned without the need of too large \(I_{33}\).

Our main purpose is to analyze the pure threshold effects as they may
arise from new physics around the electroweak scale (therefore we also
neglect the running of neutrino parameters from the electroweak to the
scale of interest). As we will see, threshold corrections to the
neutrino mass matrix can be sufficient to generate the large observed
mixings even if there is no specific mixing at the tree-level.

The tree-level mass matrix of Eq.~\eqref{eq:loop_flav} can be
transformed into the mass eigenbasis using the tree-level mixing
matrix \(\Mat{U}^{(0)}\):
\begin{equation}\label{eq:loop_mass}
m^\nul_{ij} = m^{(0)}_i \delta_{ij} + \left(m^{(0)}_i + m^{(0)}_j\right) I_{ij},
\end{equation}
where Latin indices \(i,j\) are meant to be in the mass basis and
\(I_{ij} = \sum_{\alpha\beta} I_{\alpha\beta} U^{(0)}_{\alpha i}
U^{(0)}_{\beta j}\).

In the case of exact degeneracy, \(m_1^{(0)} = m_2^{(0)} = m_3^{(0)}\),
the tree-level mixing is trivial, whereas it gets non-trivial if the
masses have different \(\CP\) parities (e.g.\ \(m_1^{(0)} = - m_2^{(0)}
= m_3^{(0)}\)) or \(\CP\) violation occurs~\cite{Wolfenstein:1981rk,
  Branco:1998bw}. Including (\(\CP\) conserving) threshold corrections,
it is possible to lift the degeneracy~\cite{Chankowski:2000fp,
  Chankowski:2001mx} without referring to \(\CP\)
phases~\cite{Branco:2014zza}. We address the question whether low-energy
threshold corrections have the power to generate significant deviations
from a given tree-level mixing with one vanishing mixing angle or even
generate the observed neutrino mixing fully radiatively in the trivial
scenario. Note that for the \(m_1^{(0)} = - m_2^{(0)} = m_3^{(0)}\)
case, there is a free rotation in the 1-3 plane left where two mixing
angles are determined from the tree-level flavor structure.

\paragraph{The \(m_1^{(0)} = - m_2^{(0)} = m_3^{(0)}\) scenario}
The assignment of different \(\CP\) eigenvalues to different masses
simplifies the situation of Eq.~\eqref{eq:loop_mass} tremendously:
\begin{equation}
\Mat{m}^\nul = m \begin{pmatrix}
1 + 2 U_{\alpha 1} U_{\beta 1} I_{\alpha\beta} & 0 &
 2 U_{\alpha 1} U_{\beta 3} I_{\alpha\beta} \\
0 & - 1 -  2 U_{\alpha 2} U_{\beta 2} I_{\alpha\beta} & 0 \\
 2 U_{\alpha 1} U_{\beta 3} I_{\alpha\beta} & 0 &
1 +  2 U_{\alpha 3} U_{\beta 3} I_{\alpha\beta}
\end{pmatrix},
\end{equation}
where summation over repeated indices is understood. The diagonalization
of the 1-3 block can be done by requiring
\begin{equation}\label{eq:13constr}
\sum_{\alpha\beta} U_{\alpha 1} U_{\beta 3} I_{\alpha\beta} = 0,
\end{equation}
which can be motivated analogously to degenerate perturbation theory in
quantum mechanics and exploits the freedom of rotation in the 1-3
plane~\cite{Chun:1999vb, Chankowski:2000fp, Chankowski:2001mx}.

In general, the threshold corrections \(I_{\alpha\beta}\) do not have
to be diagonal---however, it is intriguing to analyze whether
flavor-diagonal contributions \(I_\alpha = I_{\alpha\alpha}\)
(\(\alpha = \el,\mul,\taul\)) are sufficient to generate the desired
deviation from the degenerate pattern. Condition
Eq.~\eqref{eq:13constr} relates the third yet undetermined mixing
angle to the other two and gives a constraint on the threshold
corrections. Including all three corrections, we get
an extension of Eq.~(5.11) in \cite{Chankowski:2001mx}:
\begin{equation}\label{eq:sin13}
s_{13} = c_{23} s_{23} \frac{s_{12}}{c_{12}} \frac{I_\mul -
  I_\taul}{I_\el - s_{23}^2 I_\mul - c_{23}^2 I_\taul},
\end{equation}
where \(s_{ij}=\sin\theta_{ij}\) and \(c_{ij}=\cos\theta_{ij}\) as
defined above. This can be used to replace \(I_\el\) in terms of
\(I_\mul\), \(I_\taul\) and the mixing angles:
\begin{equation}
I_\el = s_{23}^2 I_\mul + c_{23}^2 I_\taul + \frac{c_{12} s_{23}
  c_{23}}{c_{12} s_{13}} \left(I_\mul-I_\taul\right).
\end{equation}
There are then two parameters left (\(I_\mul\) and \(I_\taul\)) to fit
the two mass squared differences. Shifting all corrections by an overall
flavor universal constant, \(I_\alpha \to \tilde I_\alpha = I_\alpha - I\), does
neither change the mixing angles nor does it affect the ratio \(\Delta
m_{31}^2 / \Delta m_{21}^2\)~\cite{Chankowski:2001mx}. Fitting
simultaneously a large \(\theta_{13}\) and the masses (\(\Delta
m_{21}^2\) and \(\Delta m_{32}^2\)) with the diagonal corrections only
is excluded with the current neutrino data for perturbative
\(I_\alpha\). Using Eq.~\eqref{eq:sin13} to eliminate \(I_\el\), the two
mass squared differences
\begin{equation}\label{eq:Deltam2corr}
\Delta m_{ab}^2 \approx \tilde m^2 \left[\left(1 + 2 U_{\alpha a} U_{\beta a}
    \tilde I_{\alpha\beta} \right)^2
 - \left(1 + 2 U_{\alpha b} U_{\beta b} \tilde I_{\alpha\beta}\right)^2\right]
\end{equation}
are then roughly of the same order instead of approximately separated by
one order of magnitude. The mass parameter \(\tilde m\) is scaled with
the shift in \(\tilde I_\alpha\): \(\tilde m = (1+ 2I) m\). To get a
larger splitting between \(\Delta m_{21}^2\) and \(\Delta m_{31}^2\) is
incompatible with the requirement~\eqref{eq:sin13}, at least for real
\(\theta_{13}\) (which constraints \(|s_{13}| \leq 1\)) and
\(\delta_\CP=0\). For a non-zero Dirac \(\CP\)-phase the expressions
become much more involved.

\begin{figure*}
\centering
\begin{minipage}{.5\textwidth}
\includegraphics[width=\textwidth]{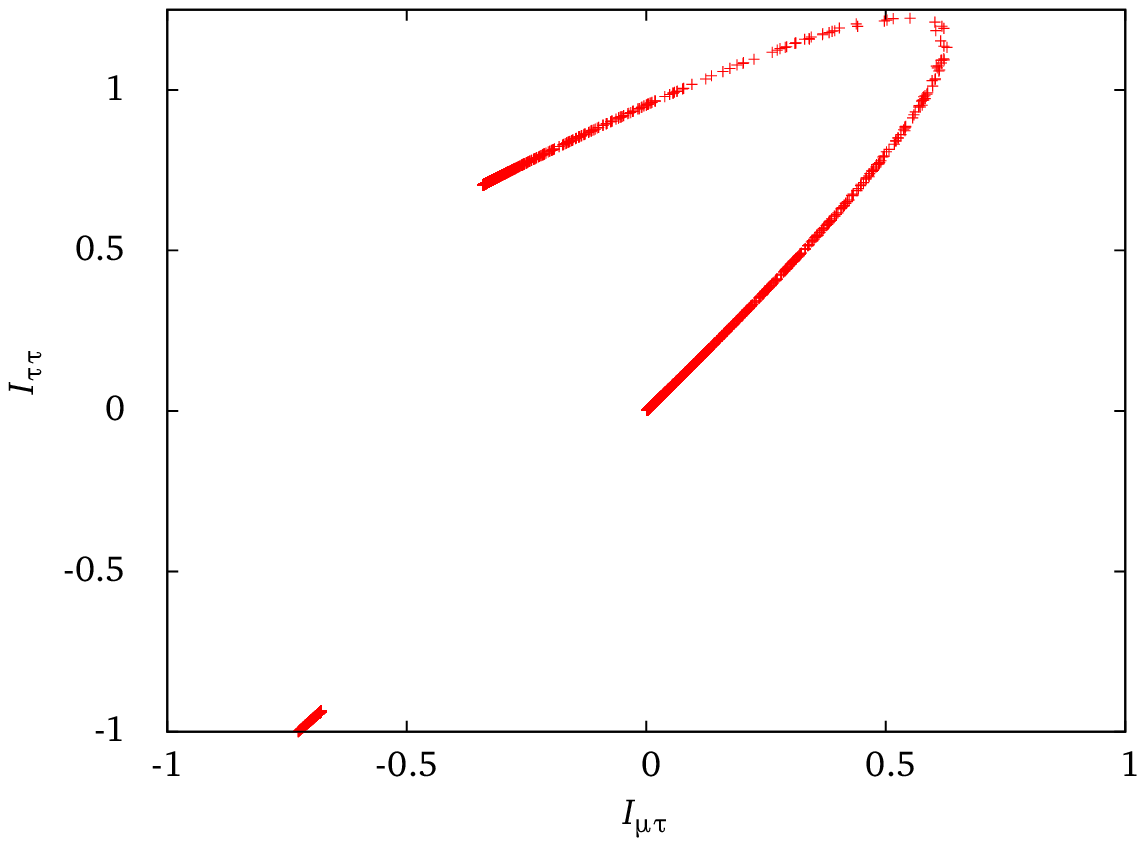}
\end{minipage}%
\begin{minipage}{.5\textwidth}
\includegraphics[width=\textwidth]{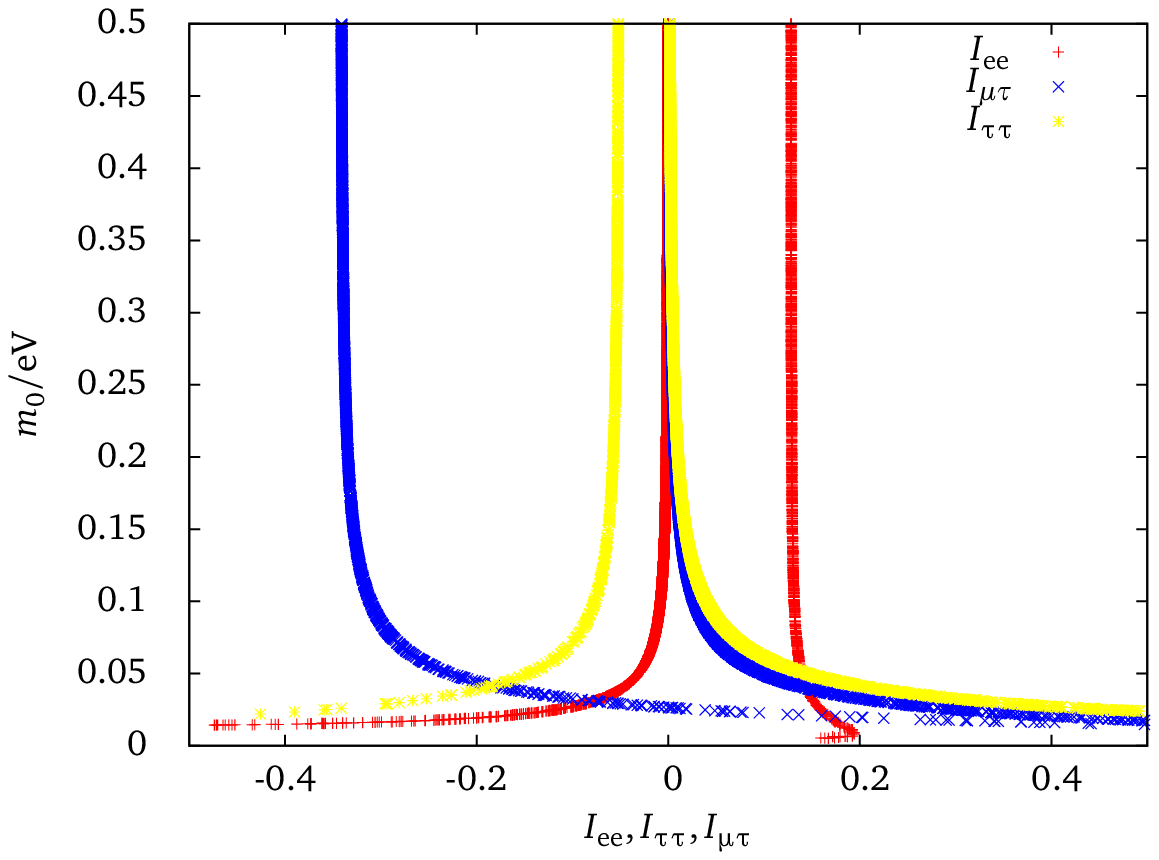}
\end{minipage}%
\caption{The allowed ranges for \(I_{\taul\taul}\) and \(I_{\mul\taul}\)
  for \(\Delta m_{31}^2\) and \(\Delta m_{21}^2\) within their
  \(1\,\sigma\) ranges (the mixing angles are taken at the central
  values). The left plot shows the dependence on the lightest neutrino
  mass \(m_0\) in the vertical direction. There is one class of
  solutions where all three non-vanishing elements of \(\Mat{I}\) are
  close to zero.} \label{fig:imutau}
\end{figure*}

Of course, the threshold corrections \(I_{\alpha\beta}\) are not
necessarily flavor diagonal. Including more contributions (with
\(\alpha\neq\beta\)) may help to get a better fit to data. However, the
new parameters are less constrained. It is interesting to remark that
out of the given relations between the mixing angles in the case where
one off-diagonal correction dominates (see~\cite{Chun:1999vb,
  Chankowski:2001mx}), the only surviving relation is the one for
dominant \(I_{\mul\taul}\) in view of recent data
(\(s_{13}=-\tan\theta_{12}\cot2\theta_{23}\)). To get the masses right,
at least one more sizable correction is necessary. Take e.g.\ \(I_\taul
\neq 0\), the condition~\eqref{eq:13constr} then gives a relation
\begin{equation}
I_\taul = - I_{\mul\taul} \frac{2 s_{13} s_{23} + \cos 2\theta_{23}
  \tan\theta_{12} / c_{23}}{c_{23} s_{13} - s_{23} \tan\theta_{12}}
\approx \frac{5}{4}\;I_{\mul\taul}.
\end{equation}
The contribution in both \(\Delta m^2\) is \(\sim m^2 I_{\mul\taul}\)
which is---for the known values of the mixing angles, where no
accidental cancellations appear---not sufficient to give rise to the
splitting \(\Delta m_{31}^2 \gg \Delta m_{21}^2\). Considering also
\(I_\el \neq 0\) helps to find solutions for not too heavy neutrinos
(\(m\approx 0.1\,\eV\)) and \(|I_\taul|,|I_{\mul\taul}| =
\mathcal{O}(10^{-2})\) via the relation
\begin{equation}\label{eq:sin13rel}
s_{13} = \frac{( I_\taul c_{23} s_{23} - I_{\mul\taul} \cos 2\theta_{23}
  ) \tan\theta_{12}}{ I_\taul c_{23}^2 + I_{\mul\taul}\sin2\theta_{23} - I_\el}.
\end{equation}
We can now see whether there are solutions for \(I_\taul\) and
\(I_{\mul\taul}\) that give the right \(\Delta m^2\) according to
Eq.~\eqref{eq:Deltam2corr}. The angles are already fixed and
Eq.~\eqref{eq:sin13rel} actually determines \(I_\el\) in terms of
\(I_\taul\) and \(I_{\mul\taul}\). Viable solutions are shown in
Fig.~\ref{fig:imutau}, where we show the allowed regions for the two
free entries of \(\Mat{I}\). The mass squared differences were fixed to
be in the \(1\,\sigma\) intervals and the mixing angles at their central
values. There is actually one class of solutions where all three
\(I_\el\), \(I_\taul\) and \(I_{\mul\taul}\) are small and get smaller
for larger values of \(m_0\). For \(m_0=0.1(0.35)\,\eV\),
we have \(I_\el = -8.15 \times 10^{-3} (-6.78 \times 10^{-4})\),
\(I_{\mul\taul} = 2.73 \times 10^{-2} (2.31 \times 10^{-3})\) and
\(I_\taul = 3.97 \times 10^{-2} (3.35 \times 10^{-3})\).

Examination of other configurations as \(m_1 = m_2 = - m_3\) or \(-m_1 =
m_2 = m_3\) are qualitatively the same and can be treated analogously.

\paragraph{The exact degenerate case: \(m_1^{(0)} = m_2^{(0)} = m_3^{(0)}\)}
Unlike the situation where one mass eigenvalue has a different sign and
there is only one freedom of rotation (in the plane where both masses
have the same sign), the exact degenerate case allows for three
arbitrary rotations. Trivially, there is no mixing matrix at the
tree-level and therefore it is interesting to figure out whether
threshold corrections have the power to generate the mixing---and lift
the degeneracy in masses.

In General, Eq.~\eqref{eq:loop_mass} is a complex symmetric
matrix, where the first term is proportional to the unit matrix:
\begin{equation}\label{eq:corr_mass}
\Mat{m}^\nul = m \;\mathds{1} + m
\begin{pmatrix}
I_{11} & I_{12} & I_{13} \\
I_{12} & I_{22} & I_{23} \\
I_{13} & I_{23} & I_{33}
\end{pmatrix},
\end{equation}
and \(m\) is the common neutrino mass. Obviously, \(\Mat{m}^\nul\) is
diagonalized by diagonalizing only the perturbation \(\Mat{I}\). Models
of exact degeneration can be motivated from \(\SO(3)\) or \(\SU(3)\)
symmetries---or finite subgroups of them. Eq.~\eqref{eq:corr_mass} is
the starting point to derive the physical mixing matrix from the
threshold corrections. To a very good approximation, the charged leptons
do not receive sizable flavor changing corrections, so the
re-diagonalization of the perturbed neutrino mass matrix gives directly
the phenomenological leptonic mixing matrix \(\Mat{U}_\upsh{PMNS}\)
observed in charged current interactions. (We work in the charged lepton
mass basis.)

The diagonal neutrino mass matrix is obtained by the use of
Eq.~\eqref{eq:standardparam}
\begin{equation} {\bar{\Mat{m}}}^\nul =
  \Mat{U}(\theta_{12},\theta_{13},\theta_{23})^\tp \; \Mat{m}^\nul \;
  \Mat{U}(\theta_{12},\theta_{13},\theta_{23}) = \Mat{U}_{12}^\tp
  \Mat{U}_{13}^\tp \Mat{U}_{23}^\tp \Mat{m}^\nul \Mat{U}_{23}
  \Mat{U}_{13} \Mat{U}_{12},
\end{equation}
which shows that in the standard parametrization \(\Mat{U}_{23}\) acts
``first'' on the full mass matrix \(\Mat{m}^\nul\). Note that for different
parametrizations, especially a different ordering of rotations, the
assignments to the measured angles are different and first have to be
re-expressed by the standard angles. Eq.~\eqref{eq:standardparam} as
mixing matrix allows to directly apply the results of
Ref.~\cite{GonzalezGarcia:2012sz} and similar results to our problem.

Experimentally, \(\theta_{23}\), the atmospheric mixing angle, is
measured to be roughly maximal \(|\theta_{23}| \approx \frac{\pi}{4}\)
with a small deviation of a few degrees.  The rotation angle in the
\(i\)-\(j\) plane can be expressed analytically in terms of the matrix
elements via
\begin{equation}\label{eq:def_theta}
\tan 2\theta_{ij} = \frac{ 2 I_{ij} }{I_{jj} - I_{ii}},
\end{equation}
where \(\theta_{ij}\) can be chosen such that
\(0\leq\theta_{ij}\leq\pi/4\) by reordering diagonals and
off-diagonals and shifting the the phase in the two-fold transformation
\[
\Mat{U_{ij}}(\theta_{ij}, \delta_{ij}) =
\begin{pmatrix}
\cos\theta_{ij} & \sin\theta_{ij} e^{-\im\delta_{ij}} \\
-\sin\theta_{ij} e^{\im\delta_{ij}} & \cos\theta_{ij}
\end{pmatrix}.
\]
Maximal mixing in the 2-3 plane means, that this sector can be
diagonalized using
\[
\Mat{U}_{23} = \begin{pmatrix}
1 & 0 & 0 \\
0 & \frac{1}{\sqrt{2}} & -\frac{1}{\sqrt{2}} \\
0 & \frac{1}{\sqrt{2}} & \frac{1}{\sqrt{2}}
\end{pmatrix},
\]
by setting \(I_{33} = I_{22}\) in Eq.~\eqref{eq:corr_mass}. For
simplicity and because it only has an influence on the resulting
eigenvalues which can be absorbed in a redefinition of the mass
parameter \(m\), we also take \(I_{23} = I_{22}\) (note that this is not
required for \(\theta_{23}=\pi/4\) and we release this requirement
later).\footnote{Actually, we have chosen an unconventional sign
  convention which corresponds rather to \(\theta_{23} = -\pi/4\)
  contrary to the conventions described beforehand. A similar result can
  be obtained for the other sign. In the given convention, \(m_3 = m\)
  and (for normal hierarchy) \(m_{1,2} < m\).} As result, we get
\begin{equation}\label{eq:first_rot}
\Mat{I}^\prime = \Mat{U}^\tp_{23} \;\Mat{I}\; \Mat{U}_{23} = \begin{pmatrix}
I_{11} & \frac{I_{12}+I_{13}}{\sqrt{2}} & -
\frac{I_{12}-I_{13}}{\sqrt{2}} \\
\frac{I_{12}+I_{13}}{\sqrt{2}} & 2 I_{22} & 0 \\
-\frac{I_{12}-I_{13}}{\sqrt{2}} & 0 & 0
\end{pmatrix}.
\end{equation}

Since a few years ago, the measured value of \(\theta_{13}\) was
comparable to zero, which suggests furthermore the approximation
\(I_{13} \approx I_{12}\) such that the last rotation is given by
\begin{equation}\label{eq:theta12}
\theta_{12} \approx \frac{1}{2} \arctan\left(\frac{2\sqrt{2} I_{12}}{2I_{22} -
  I_{11}}\right)
\end{equation}
and the eigenvalues of \(\Mat{m}^\nul\) are \(m_3 = m\) and
\begin{equation}\label{eq:masses1+2}
m_{1,2} = m \left\lbrace
1 + \frac{1}{2} \left[
I_{11} + 2 I_{22} \pm \left(I_{11} - 2 I_{22}\right)
\sqrt{\frac{(I_{11}-2 I_{22})^2 + 8 I_{12}^2}{(I_{11}-2 I_{22})^2}}
\right]\right\rbrace.
\end{equation}
The mass squared differences calculated from the masses in
\eqref{eq:masses1+2} can be obtained as
\begin{equation}\label{eq:delta12}
\begin{aligned}
\Delta m_{21}^2 &= m^2 \left(2 + I_{11} + 2 I_{22}\right)
\left(2 I_{22} - I_{11}\right) \sqrt{ 1 + \frac{8 I_{12}^2}{\left(I_{11}
      - 2 I_{22}\right)^2} }, \\
\Delta m_{31}^2 &= -\frac{m^2}{2} \Bigg[ \left(
    I_{11}(I_{11} + 2) - I_{22}(I_{22} + 1) \right)
  \sqrt{\frac{ \left(I_{11} - 2 I_{22} \right)^2 + 8 I_{12}^2}{(I_{11}-2
      I_{22})^2}} \\
&\qquad\qquad\;
+ I_{11} \left( I_{11} + 2 \right) + I_{22} \left( I_{22} + 1\right) +
4 I_{12}^2 \Bigg].
\end{aligned}
\end{equation}
Altogether, there are four free parameters left (\(m\), \(I_{11}\),
\(I_{22}\) and \(I_{12}\)) required for fitting three masses and one
mixing angle (\(\theta_{12}\)).
The other two mixing angles were set to
phenomenologically motivated distinct values (\(\theta_{13}=0\) and
\(\theta_{23}=-\pi/4\)) and shall receive small corrections in the following.

Up to now, we have set the third mixing angle to zero, which is
disfavored by current experimental data. Nevertheless, we want to take
the observed pattern in the quantum corrections as starting point to
evaluate deviations from that by assigning deviations to the two
restrictions that were set explicitly:
\begin{equation}\label{eq:defdevia}
\begin{aligned}
I_{33} &= I_{22} + \varepsilon,\\
I_{13} &= I_{12} + \delta,
\end{aligned}
\end{equation}
with \(\varepsilon,\delta\) parametrizing the deviations. As we will
see, \(\varepsilon\) and \(\delta\) do not necessarily have to be small
compared to \(I_{12}\) and \(I_{22}\). Especially the deviation in
\(I_{33}\) has to be of the same order as \(I_{22}\). In that way, we
now have six parameters (\(I_{11}\), \(I_{22}\), \(I_{12}\), \(I_{23}\),
\(\delta\) and \(\varepsilon\)) to completely fit three masses and three
mixing angles, where \(\theta_{23}\) is expected to be close to
\(\frac{\pi}{4}\) and \(\theta_{13}\) small. We assign the
``unperturbed'' mass parameter to be \(m=m_0\), any flavor-universal
contribution in the threshold corrections can be simply added as a shift
in the diagonals: \(\tilde I_\alpha = I_\alpha - I_0\) for
\(\alpha=\el,\mul,\taul\).  The relation between the observed masses and
the threshold corrections is given by
\begin{equation}\label{eq:fullcase}
\begin{aligned}
& \begin{pmatrix}
m_0 &&\\ &\sqrt{m_0^2+\Delta m_{21}^2}& \\ &&\sqrt{m_0^2+\Delta m_{31}^2}
\end{pmatrix}
\\&\qquad = m_0\, \Mat{U}(\theta_{12},\theta_{13},\theta_{23})^\tp
\begin{pmatrix}
1 + I_{11} & I_{12} & I_{12} + \delta \\
I_{12} & 1 + I_{22} & I_{23} \\
I_{12} + \delta & I_{23} & 1 + I_{22} + \varepsilon
\end{pmatrix}
\Mat{U}(\theta_{12},\theta_{13},\theta_{23}),
\end{aligned}
\end{equation}
where the mixing angles \(\theta_{ij}\) and mass squared differences are
fixed by experiment. For the numerical values, we refer to the global
fit from the \(\nu\)-fit collaboration~\cite{GonzalezGarcia:2012sz},
where we now focus on the central value as a proof of principle:
\[
\begin{aligned}
\theta_{12} \approx 31.8^\circ \;,\; \theta_{13} &\approx 8.5^\circ \;,\;
\theta_{23} \approx 39.2^\circ \;,\\
\Delta m_{21}^2 \approx 7.5 \times 10^{-5}\,\eV^2 \;&,\;
\Delta m_{31}^2 \approx 2.458 \times 10^{-3} \,\eV^2.
\end{aligned}
\]
The lightest neutrino mass \(m_0\) is in principle a free parameter that
will be tested by future neutrino mass experiments. Even if \(m_0\) is
not large enough to be directly measured, we show that quantum
corrections still can lift the degenerate mass pattern sufficiently.
Therefore we compare the two cases where either a positive direct
determination is to be expected (\(m_0=0.35\,\eV\)) and the
cosmologically favored (\(m_0=0.1\,\eV\)). The second value is still
compatible within about \(1\,\sigma\) with the \(95\,\%\) upper bound on
\(\sum m_\nu\) and the nonzero value from galaxy clustering and
lensing~\cite{Ade:2013zuv, Battye:2013xqa}.

\begin{table*}[b,t]
\caption{Values of the threshold corrections needed to obtain the
  observed mixing angles and mass splitting for a common neutrino mass
  of \(0.1\,\eV\) and \(0.35\,\eV\).}\label{tab:thresholds}
\centering
\begin{tabular}{ccc}
\toprule
& \(m_0=0.1\,\eV\) & \(m_0=0.35\,\eV\) \\
\midrule
\(I_{11}\) & \(3.54\times 10^{-3}\) & \(3.00 \times 10^{-4}\) \\
\(I_{12}\) & \(1.19 \times 10^{-2}\) & \(1.02 \times 10^{-3}\) \\
\(I_{22}\) & \(4.67 \times 10^{-2}\) & \(4.01 \times 10^{-3}\) \\
\(I_{23}\) & \(5.43 \times 10^{-2}\) & \(4.67 \times 10^{-3}\) \\
\(\varepsilon\) & \(2.28 \times 10^{-2}\) & \(1.96 \times 10^{-2}\) \\
\(\delta\) & \(6.73 \times 10^{-5}\) & \(1.56 \times 10^{-5}\)\\
\(\Mat{I}\) & \(\qquad\begin{pmatrix}
  0.354 & 1.19 & 1.20 \\
  1.19 & 4.67 & 5.43 \\
  1.20 & 5.43 & 6.96
  \end{pmatrix} \times 10^{-2}\)
&
\(\qquad\begin{pmatrix}
  0.300 & 1.02 & 1.03 \\
  1.02 & 4.01 & 4.67 \\
  1.03 & 4.67 & 5.97
  \end{pmatrix} \times 10^{-3}\) \\
\bottomrule
\end{tabular}
\end{table*}

The results are shown in Tab.~\ref{tab:thresholds} where we compare the
two scenarios of a cosmologically inspired quasi-degenerate spectrum and
a strong degeneracy as would follow from a KATRIN neutrino mass
measurement. It is amusing to see that the entries of \(\Mat{I}\) are in
both cases of the size of a typical radiative correction
\(\lesssim\mathcal{O}(1/100)\) (note that we only wanted to generate
tiny deviations from the degenerate pattern in a regime where the
physical masses are only slightly non-degenerate) and show a hierarchy
as \(1<2<3\) for labeling the generations. This observation can be used
in any new physics model with flavor changing low-energy threshold
corrections. An interesting model is presented in the next section,
where we apply the concept of radiative generation of neutrino mixing
and mass differences to an extension of the Minimal Supersymmetric
Standard Model (MSSM). The complete contributions in the MSSM were
already calculated and rather lengthy~\cite{Chankowski:2001hx}. We are
only interested in the contributions from Supersymmetry breaking,
therefore we work with a reduced set of threshold corrections. Although
we do not refer to any flavor symmetry, there are models with a
high-energy non-abelian symmetry as \(A_4\) with degenerate masses where
the mass splitting and corrections to the mixing angles also happen
radiatively~\cite{Babu:2002dz, Morisi:2013qna}.

\begin{figure*}[tb]
\begin{minipage}{0.5\textwidth}
\includegraphics[width=\textwidth]{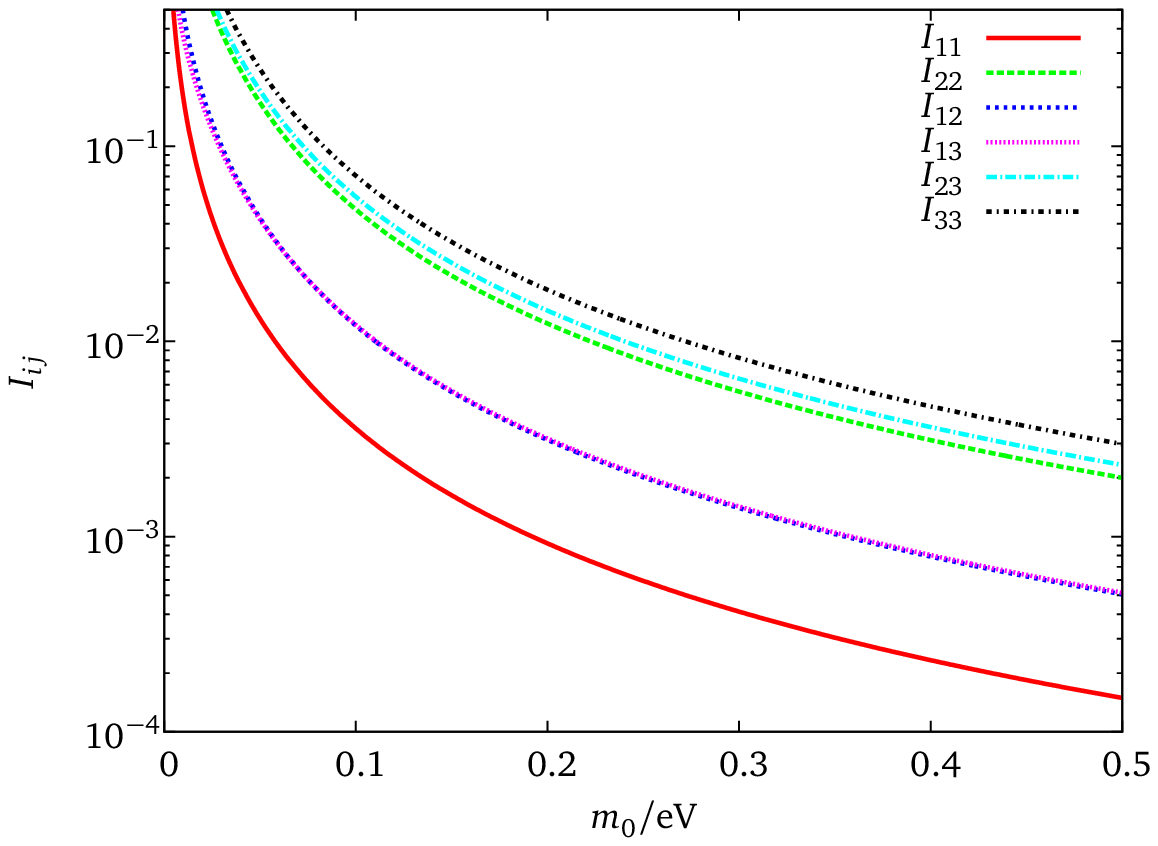}
\end{minipage}%
\begin{minipage}{0.5\textwidth}
\includegraphics[width=\textwidth]{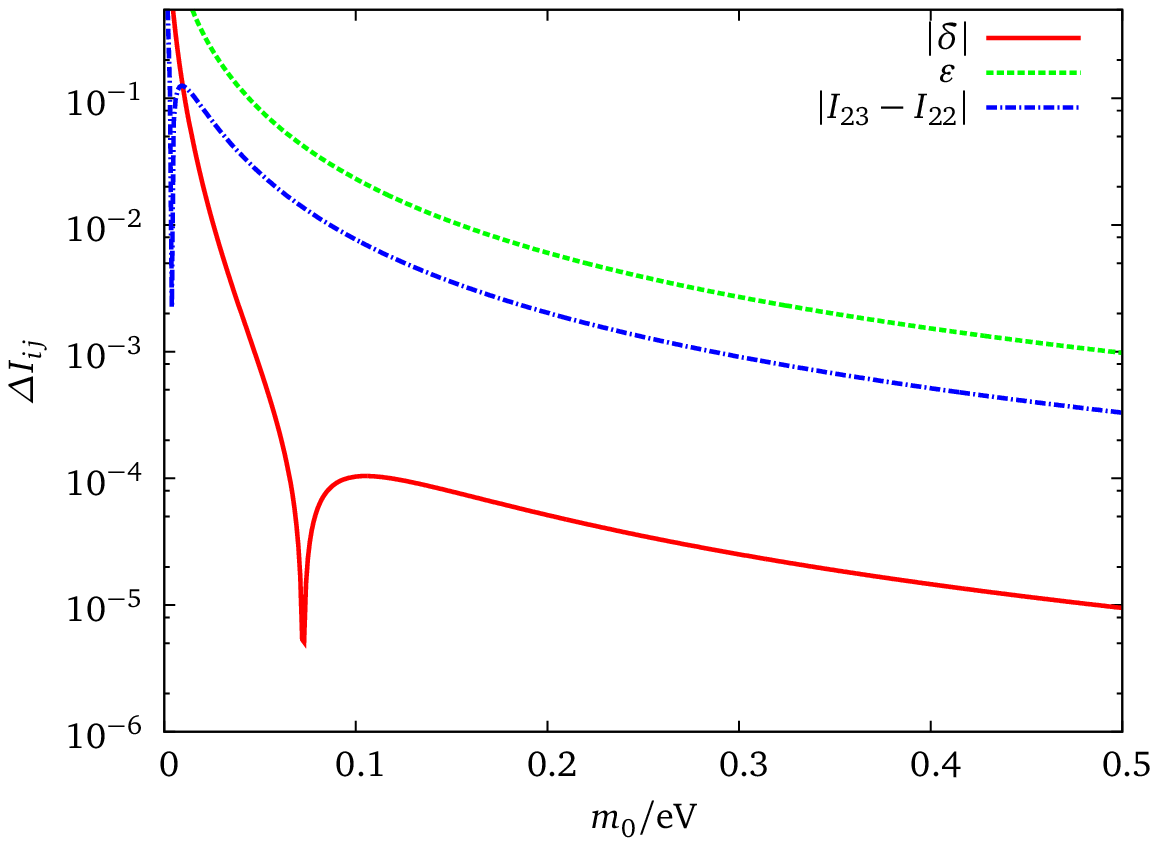}
\end{minipage}\\
\begin{minipage}{0.5\textwidth}
\includegraphics[width=\textwidth]{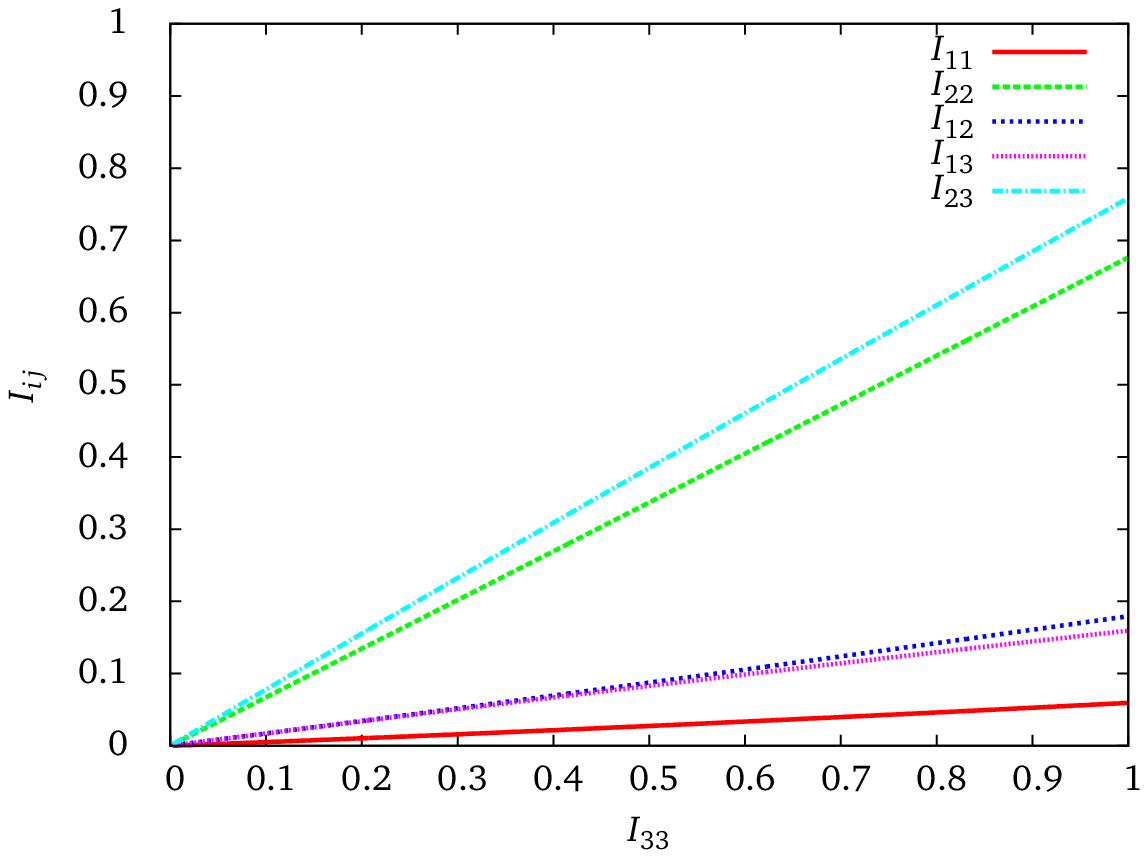}
\end{minipage}%
\begin{minipage}{0.5\textwidth}
\includegraphics[width=\textwidth]{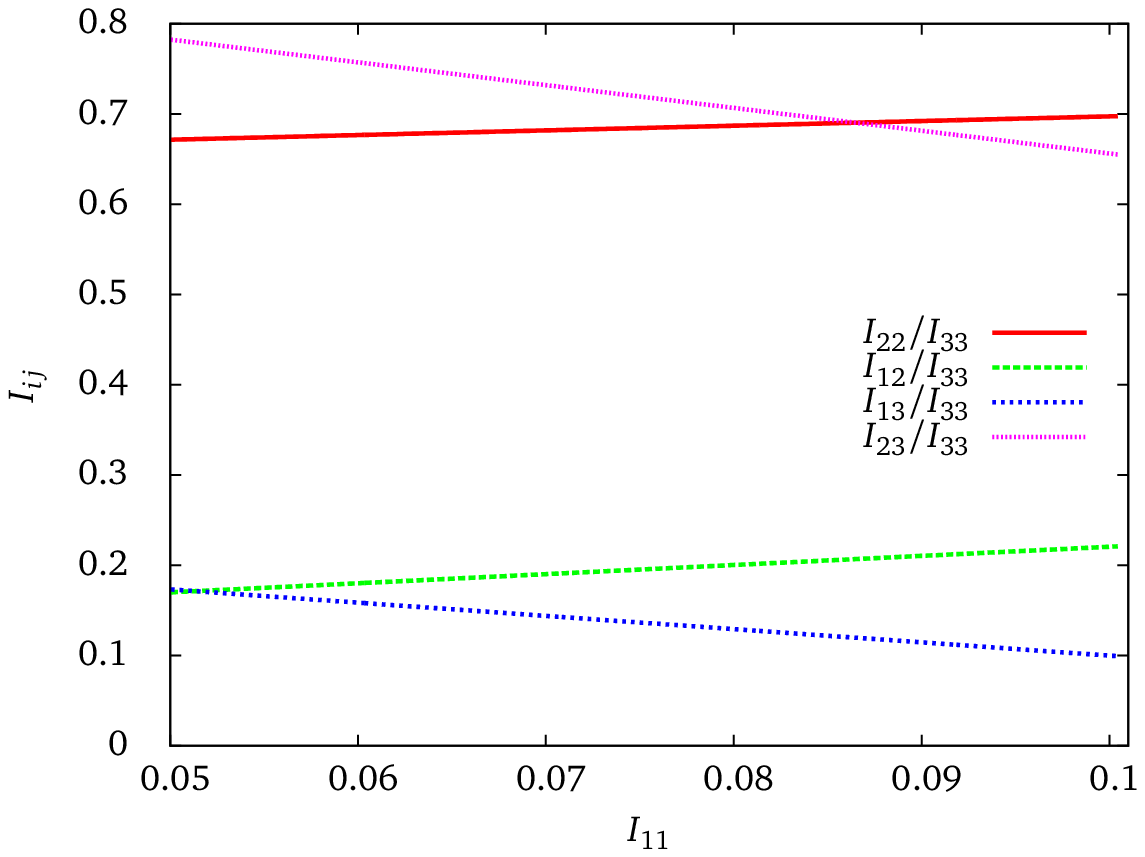}
\end{minipage}
\caption{Graphical representation of the individual corrections. The
  first row shows the dependence on the absolute neutrino mass scale:
  the larger \(m_0\) the smaller the corrections can be. The upper right
  plot shows the deviations from equal values: \(\delta =
  I_{13}-I_{12}\) and \(\varepsilon = I_{33} - I_{22}\) es defined in
  Eq.~\eqref{eq:defdevia}. The lower line shows the interplay of the
  individual \(I_{jk}\) compared to \(I_{33}\), similar plots can be done for other
  combinations. The lower right plot shows the \(I_{jk}\) normalized to
  the largest contribution \(I_{33}\).}\label{fig:thresholds}
\end{figure*}

A crucial point in the discussion is the behavior of the generic
threshold corrections with the lightest neutrino mass. In case the
overall mass scale \(m_0\) drops below \(0.1\,\eV\), the spectrum looses
the degenerate property which is expressed in values \(I_{\alpha\beta}
\simeq 0.1\) as can be seen in Fig.~\ref{fig:thresholds}. Corrections
are needed that are not of the size of typical perturbative
corrections. The hierarchical regime (\(m_0 \ll 0.1\,\eV\)) needs a
special kind of flavor symmetry breaking where the degenerate patterns
only needs a symmetry that guarantees equal masses. For a given symmetry
breaking chain, the hierarchy can be exploited to construct the mixing
matrix out of the mass ratios~\cite{Hollik:2014jda}.

\section{A viable example: threshold corrections in the \(\nul\)MSSM without minimal flavor violation}
\label{sec:nuMSSM}
The MSSM is known to come along with many new sources of flavor
violation in general. Incorporating the type-I seesaw mechanism to
arrive at the dimension five operator \eqref{eq:WeinbergOp} allows for
an arbitrary flavor pattern arising at the loop-level even if some
symmetry preserves flavor blind (and therefore degenerate) patterns at
the tree-level. We will denote the seesaw-extension of the MSSM which is
described in the following by \(\nul\)MSSM.

Let us, for aesthetic reasons, introduce the same number of right-handed
neutrino superfields as there are left-handed \(\SU(2)\) doublets.
Right-handed neutrinos are singlets under the SM gauge group which
allows to acquire a Majorana mass term by some not further specified
mechanism at the high scale \(M_\Rr\).

The relevant part for neutrino physics of the \(\nul\)MSSM is given by
the superpotential
\begin{equation}\label{eq:nuMSSM}
\begin{aligned}
\mathcal{W} \;\supset\; &\mu H_1 \cdot H_2
 + Y^\nul_{ij}\; H_2 \cdot L_{\Ll,i} N_{\Rr,j} \\
& - Y^\ell_{ij}\; H_1 \cdot L_{\Ll,i} E_{\Rr,j}
 + \frac{1}{2} M^\Rr_{ij} N_{\Rr,i} N_{\Rr,j},
\end{aligned}
\end{equation}
where we have the two Higgs
\(\SU(2)\) doublets \(H_1=(h_1^0,h_1^-)\) and \(H_2=(h_2^+,h_2^0)\),
and the dot product denotes \(\SU(2)\)-invariant multiplication. The
doublet of left-handed leptons is written as \(L_\Ll=(N_\Ll,E_\Ll)\),
where capital letters denote chiral superfields \(F=\lbrace\tilde
f,f\rbrace\) and the right-handed matter fields are contained in
the left-chiral multiplets
\(E_\Rr=\lbrace\tilde\ell^*_\Rr,\ell_\Rr^c\rbrace\) and
\(N_\Rr=\lbrace\tilde\nu^*_\Rr,\nu^c_\Rr\rbrace\). Generation indices
\(i,j\) are used in an obvious manner, \(f^c\) is the charge conjugated
fermion component.

To break Supersymmetry (SUSY) softly, we introduce the following potential terms
\begin{equation}\label{eq:Vsoft}
V_\mathrm{soft}^{\tilde\nu} =
\left(\Mat{m}_{\tilde \Ll}^2\right)_{ij} \tilde{\nu}_{\Ll,i}^*\tilde\nu_{\Ll,j}
+
\left(\Mat{m}_{\tilde \Rr}^2\right)_{ij} \tilde\nu_{\Rr,i}\tilde{\nu}^*_{\Rr,j}
+
\left(
A^\nul_{ij}\; h_2^0\, \tilde\nu_{\Ll,i}\tilde{\nu}_{\Rr,j}^*
+
\left(\Mat{B}^2\right)_{ij} \tilde{\nu}_{\Rr,i}^*\tilde{\nu}_{\Rr,j}^*
+ \hc
\right).
\end{equation}
The neutrino \(B\) term is written in a way that suggests no connection
to \(M_\Rr\) although it can be seen as a ``Majorana-like'' soft
breaking mass term (therefore denoted here as \(\Mat{B}^2\)). The usual way to
write it down in the literature is rather \(\Mat{B}^2 = b\,\Mat{M}_\Rr\) with \(b\)
being a parameter of the SUSY scale, see
e.g. \cite{Farzan:2004cm,Dedes:2007ef,Heinemeyer:2014hka}.

In general, \(\Mat{A}^\nul\) as well as \(\Mat{B}^2\) are arbitrary matrices in
flavor space (\(\Mat{B}^2\) is symmetric). Because flavor-off-diagonal
entries in the soft breaking mass \(\Mat{m}_{\tilde\Ll}^2\) easily lead to large
FCNC processes in charged lepton physics, we take this contribution
flavor blind as well as \(\Mat{m}_{\tilde\Rr}^2\):
\[\Mat{m}_{\tilde\Ll}^2 = \Mat{m}_{\tilde\Rr}^2 = M_\upsh{SUSY}^2\,\mathds{1}.\]

Without loss of generality, we work in a basis where the charged lepton
Yukawa coupling as well as the right-handed Majorana mass is
diagonal. The neutrino Yukawa coupling can then be expressed in terms of
the right-handed masses \(M^\Rr_i\), the light neutrino masses
\(m^\nul_i = v_\uq^2 \kappa_i\) and the PMNS matrix \cite{Casas:2001sr}:
\begin{equation}\label{eq:Casas-Ibarra}
\Mat{Y}_\nul = \sqrt{\Mat{M}_\Rr} \mathcal{R} \sqrt{\Mat{\kappa}} \Mat{U}_\upsh{PMNS}^\dag,
\end{equation}
with \(\mathcal{R}\) being an (arbitrary) complex orthogonal matrix. The
Matrices \(\sqrt{\Mat{M}_\Rr}\) and \(\sqrt{\Mat{\kappa}}\) are diagonal
matrices of the heavy and light masses, respectively. The right-handed
Majorana mass scale \(M_\Rr\) is \emph{a priori} not constrained, where
limits can be set from leptogenesis~\cite{Davidson:2008bu}. To get
\(\mathcal{O}(1)\) neutrino Yukawa couplings, we set \(M_\Rr = 10^{14}\,\GeV\).

Unfortunately, Eq.~\eqref{eq:Casas-Ibarra} allows for random flavor
structures in \(\Mat{Y}_\nul\) due to \(\mathcal{R}\) which actually do
not affect the tree-level mass and mixing formulae. Without any
restrictions on \(\Mat{Y}_\nul\), any prediction on the flavor mixing
behavior of SUSY threshold corrections would be useless, because not
only the combination \(\Mat{Y}_\nul^\tp \Mat{M}_\Rr^{-1} \Mat{Y}_\nul\) will
appear but also \(\Mat{A}^\nul \Mat{Y}_\nul\) and \(\Mat{M}_\Rr
\Mat{Y}_\nul\).

Remember that we wanted to explain deviations from the degenerate
neutrino mass pattern. Integrating out the heavy superfields in
Eq.~\eqref{eq:nuMSSM} brings us to the effective operator of
Eq.~\eqref{eq:WeinbergOp} and yields a light neutrino mass matrix of the
form
\begin{equation}\label{eq:numass}
\Mat{m}_\nul^{(0)} = - v_\uq^2 \Mat{Y}_\nul^\tp \, \Mat{M}_\Rr^{-1} \Mat{Y}_\nul + \mathcal{O}(v_\uq^4/M_\Rr^3).
\end{equation}
To get exact degeneracy in \(\Mat{m}_\nul^{(0)}\) there has to be some
conspiracy at work that adjusts \(\Mat{Y}_\nul\) in a way to cope with
any non-degenerate pattern in \(\Mat{M}_\Rr\). Avoiding any
conspiracies, we assume (conspire?) the Yukawa coupling as well as the
right-handed Majorana mass to be flavor blind: \(\Mat{Y}_\nul =
y_\nul\,\mathds{1}\) and \(\Mat{M}_\Rr =
m_\Rr\,\mathds{1}\). Other popular choices like \[\Mat{Y}_\nul
  = \begin{pmatrix} 1&0&0\\0&0&1\\0&1&0\end{pmatrix}\] as favored by
  discrete flavor symmetries do not alter the qualitative features of
  the results. The degenerate neutrino mass is then given by \(m =
v_\uq^2 y_\nul^2 / m_\Rr\).

Calculation of the SUSY threshold corrections is done by evaluating the
neutrino self-energies including superpartners~\cite{Dedes:2007ef}:
\begin{equation}\label{eq:SUSY1loop}
\left(\Mat{m}_\nul^\mathrm{1-loop}\right)_{ij} =
\left(\Mat{m}^{(0)}_\nul\right)_{ij} +
 \Re\left[ \Sigma^{(\nul), \upsh{S}}_{ij}
   + \frac{m^{(0)}_{\nul_{i}}}{2} \Sigma^{(\nul), \upsh{V}}_{ij}
   + \frac{m^{(0)}_{\nul_{j}}}{2} \Sigma^{(\nul), \upsh{V}}_{ji}
\right],
\end{equation}
with the decomposition of the neutrino self-energy
\begin{equation}\label{eq:SigmaNu}
\Sigma^{(\nul)}_{ij} (p) = \Sigma^{(\nul),\upsh{S}}_{ij} (p^2) P_\Ll + {\Sigma^{(\nul),\upsh{S}}_{ij}}^* (p^2) P_\Rr
+ \slashed p \left[{\Sigma^{(\nul),\upsh{V}}_{ij}} (p^2) P_\Ll + {\Sigma^{(\nul),\upsh{V}}_{ij}}^* (p^2) P_\Rr\right].
\end{equation}
For Majorana neutrinos, the self-energy is flavor symmetric
(\(\Sigma_{ij} = \Sigma_{ji}\) and the coefficients in front of the left
and right projectors (\(P_\Ll\) and \(P_\Rr\) respectively) are related via
complex conjugation. We evaluate the self-energies at \(p^2=0\) since
the sparticles in the loop are superheavy compared to neutrinos.

The flavor changing scalar and vectorial parts of the neutrino
self-energies according to Eq.~\eqref{eq:SigmaNu} can be easily
calculated---although the mixing matrices are only to be determined
numerically:
\begin{subequations}
\begin{align}
  \left(\Mat{\Sigma}^{(\nul),\upsh{S}}\right)_{ij} &=
  \frac{1}{4(2\pi)^2} B_0(m_{\tilde \chi_k^0}, m_{\tilde \nu_s})
  m_{\tilde\chi_k^0} \left(\frac{-i}{\sqrt{2}}\right)^2 \left(g_2
    Z^\upsh{N}_{2k} - g_1 Z^\upsh{N}_{1k}\right)^2
  \mathcal{Z}^{\tilde\nul*}_{i's} \mathcal{Z}^{\tilde\nul*}_{j's}
  (\Mat{U}_\upsh{PMNS})_{i'i} (\Mat{U}_\upsh{PMNS})_{j'j}, \\
  \left(\Mat{\Sigma}^{(\nul),\upsh{V}}\right)_{ij} &=
  \frac{1}{4(2\pi)^2} B_1(m_{\tilde \chi_k^0}, m_{\tilde \nu_s})
  \left(\frac{-i}{\sqrt{2}}\right)^2 \left|g_2 Z^\upsh{N}_{2k} - g_1
    Z^\upsh{N}_{1k}\right|^2 \mathcal{Z}^{\tilde\nul}_{i's}
  \mathcal{Z}^{\tilde\nul*}_{j's} (\Mat{U}^*_\upsh{PMNS})_{i'i}
  (\Mat{U}_\upsh{PMNS})_{j'j},
\label{eq:SelfEnLL}
\end{align}
\end{subequations}
where summation over repeated indices is understood. The unrenormalized
neutrino mixing matrix which occurs in the vertex is denoted by
\(\Mat{U}_\upsh{PMNS}\), \(\mathcal{Z}^{\tilde\nul}\) is the slightly
modified sneutrino mixing matrix described in the appendix~\ref{app:FR},
\(\Mat{Z}^\upsh{N}\) is the neutralino mixing matrix, \(B_0\) and
\(B_1\) the emerging loop functions. The same expressions were found by
the authors of~\cite{Dedes:2007ef} as well with the conventions of
App.~\ref{app:FR}.

We are restricting ourselves to only \(\Mat{A}^\nul\) having an arbitrary
flavor structure---and determine this structure by comparing the result
of Eq.~\eqref{eq:SUSY1loop} by virtue of SUSY corrections to the
solution of Eq.~\eqref{eq:fullcase} where the size of the threshold
corrections can be obtained as
\( m\, I_{ij} = \left(\Mat{m}_\nul^\mathrm{1-loop}\right)_{ij} - m \delta_{ij}\)
with \(m\) being the degenerate mass at tree-level. The dependence on
the remaining SUSY parameters is quite mild in any respect. For the
analysis presented here we vary the values of the following variables
randomly in the given intervals:
\begin{equation}\label{eq:scatter}
\begin{aligned}
M_\upsh{SUSY} &\in [ 500, 5000 ]\,\GeV, \\
M_1 &\in [ 0.3, 3 ]\, M_\upsh{SUSY}, \\
M_2 &\in [ 1, 5 ]\, M_\upsh{SUSY}, \\
\mu &\in [ -15, 15 ]\,\TeV,\\
\tan\beta &\in [ 10, 60 ].
\end{aligned}
\end{equation}

\begin{figure*}[tb]
\begin{minipage}{.5\textwidth}
\includegraphics[width=\textwidth]{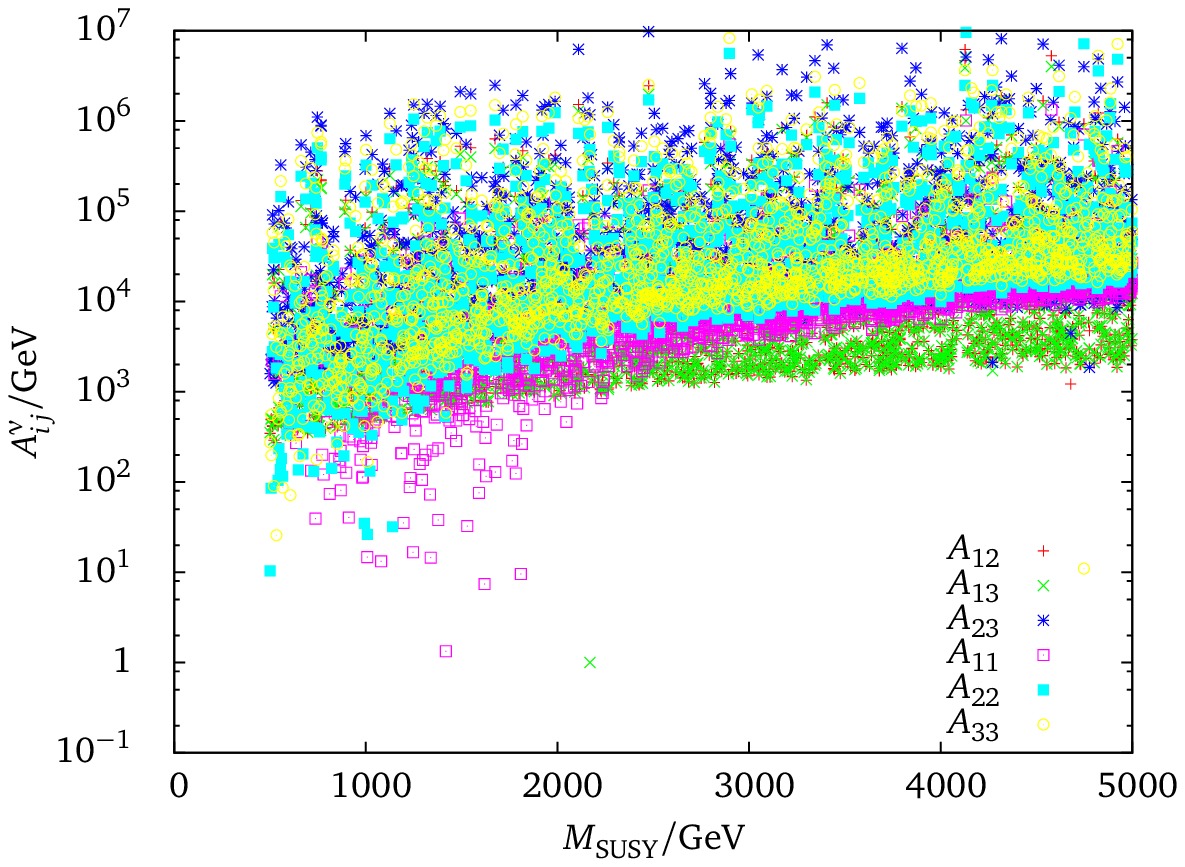}
\end{minipage}%
\begin{minipage}{.5\textwidth}
\includegraphics[width=\textwidth]{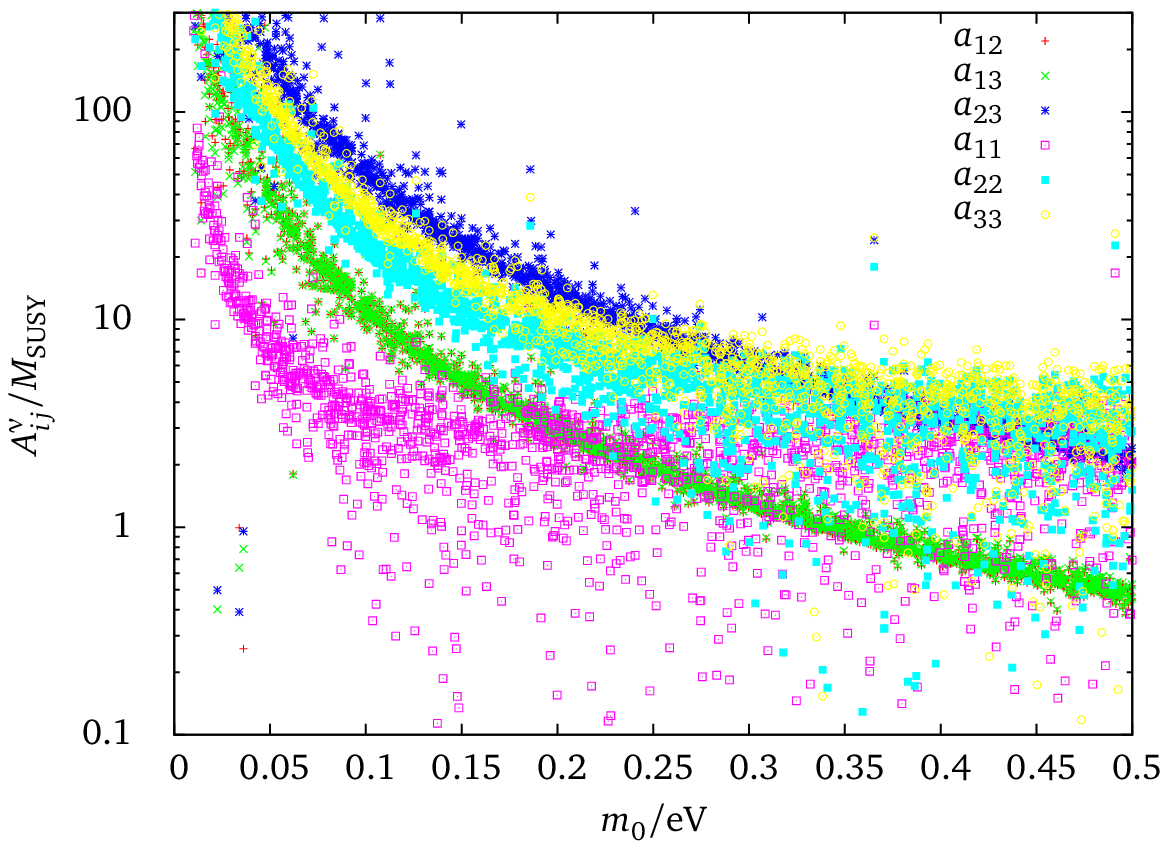}
\end{minipage}%
\caption{The left plot shows values of \(\Mat{A}^\nul\) for the
    variation of parameters specified in~\eqref{eq:scatter}. The lightest
    neutrino mass \(m_0\) was chosen in the regime plotted on the right
    side, where we rescaled all trilinear soft breaking couplings with
    the SUSY scale, \(a_{ij} = A^\nul_{ij}/M_\upsh{SUSY}\). The values
    of \(a_{12}\) and \(a_{13}\) are roughly the same since they differ
    only by a small parameter as described in Sec.~\ref{sec:general}.}
\label{fig:Anu}
\end{figure*}

As expected, for low values of the absolute neutrino mass \(m_0\) where
the deviation from the degenerate pattern is large, the SUSY threshold
corrections measured in the values of \(\Mat{A}^\nul\) have to be large as
shown in Fig.~\ref{fig:Anu} where we plotted the ratio \(a_{ij} =
A^\nul_{ij} / M_\upsh{SUSY}\). The left-hand side of Fig.~\ref{fig:Anu}
compared to the right-hand side shows that that basically this ratio is
the parameter which drives the corrections and has the same shape as the
\(I_{ij}\) dependent on \(m_0\) where the size of the \(A^\nul_{ij}\)
depending on the SUSY scale also is sensitive to the parameters of the
theory.

\begin{figure*}[tb]
\begin{minipage}{.5\textwidth}
\includegraphics[width=\textwidth]{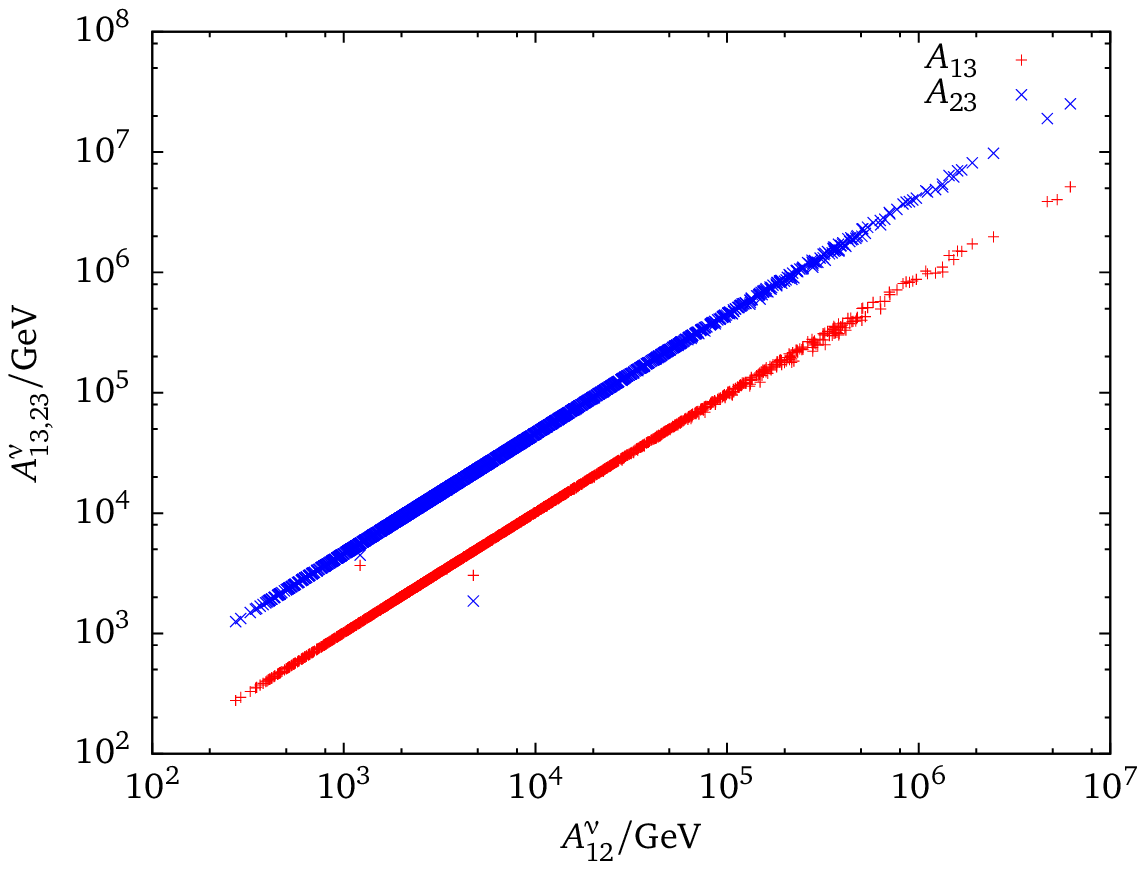}
\end{minipage}%
\begin{minipage}{.5\textwidth}
\includegraphics[width=\textwidth]{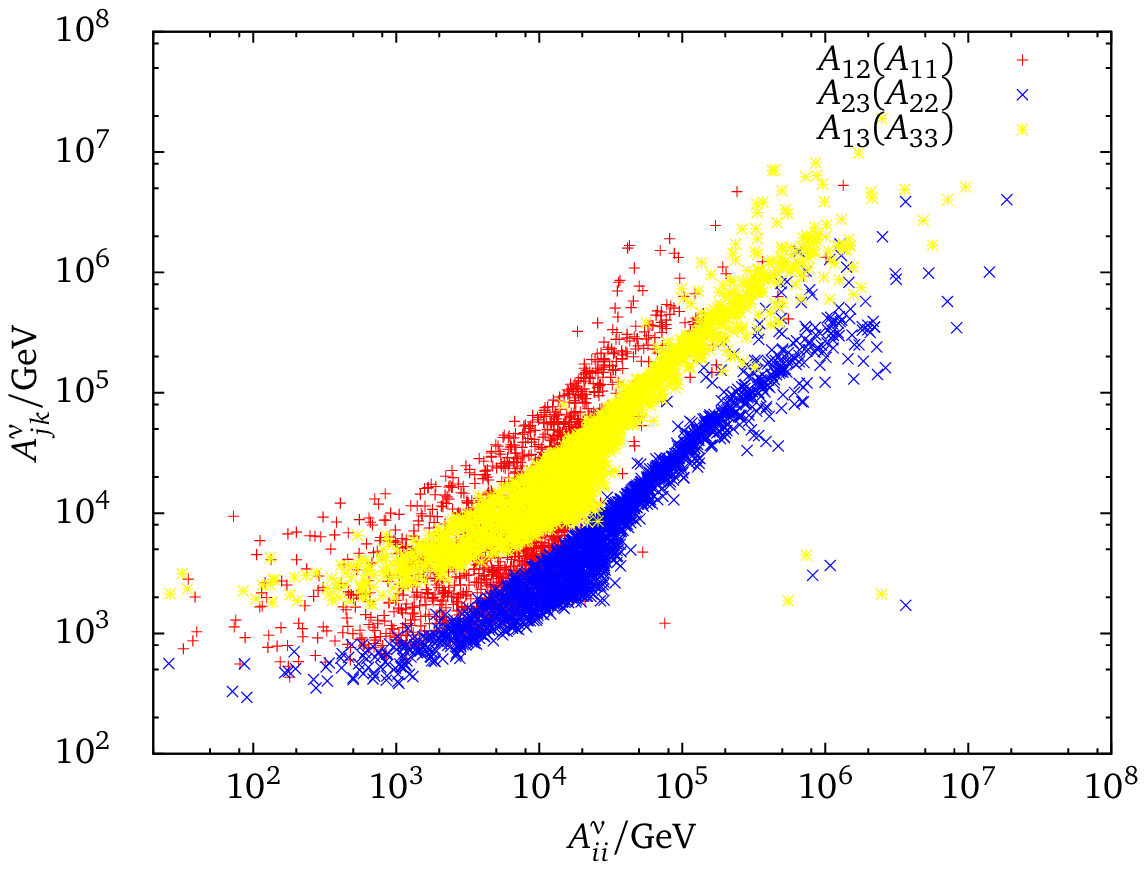}
\end{minipage}%
\caption{Correlations between the different elements of
  \(\Mat{A}^\nul\). The left plot shows the off-diagonals with respect
  to \(A^\nul_{12}\) where the right shows the correlation with the
  diagonal entries.}
\label{fig:correlA}
\end{figure*}

The correlations between the \(I_{ij}\) elaborated in
Sec.~\ref{sec:general} also get reflected to the trilinear SUSY breaking
terms shown in Fig.~\ref{fig:correlA}. Especially the off-diagonals are
very strongly correlated where we only expected this for \(A^\nul_{12}\)
and \(A^\nul_{13}\). The least correlation can be seen between
\(A_{12}\) and \(A_{11}\) which correspond to \(I_{12}\) and \(I_{11}\)
that were seen to be the most independent contributions.

\section{Conclusions}
We have discussed degenerate neutrino masses at the tree-level and
showed how threshold corrections to the masses affect the mixing. The
cosmologically favored value for the absolute neutrino mass is rather
at the edge of what is usually called quasi-degenerate mass spectrum. In
contrast, any direct measurement like from tritium decay would
immediately pose a highly degenerate spectrum. Both degenerate cases
imply that small threshold corrections arising at some scale between the
electroweak and the scale of any UV complete theory are sufficient to
generate both the neutrino mass spectrum as well as the mixing angles
radiatively. For the general discussion, we have not fixed any model to
account for neutrino masses and even neglected the RGE running down to
the electroweak scale. RGE corrections are known to be important
altering both mixing angles and mass differences considerably. Several
degenerate patterns, however, are known to preserve specific mixing
patterns.

We re-examined the more general possibility of neutrino mass eigenstates
having different \(\CP\) eigenvalues (when \(\CP\) is conserved) which
lead to a simplified description of threshold corrections. In this
scenario, there exist non-trivial tree-level mixings and the task is to
determine deviations from that. It is, however, impossible to
simultaneously accommodate for both mass splittings and a sizable third
mixing angle with only flavor-diagonal threshold corrections. We have to
include at least one off-diagonal, \(I_{\mu\tau}\), which is indeed
sufficient to reproduce the masses and mixings in the interplay with the
diagonal \(I_\el\) and \(I_\taul\).

For the case of a trivial tree-level mixing with three exactly
degenerate masses, we derived phenomenologically out of the observed
neutrino mixing patterns (e.g nearly maximal \(\theta_{23}\) whereas
small \(\theta_{13}\)) constraints and correlations on the threshold
corrections \(I_{ij}\) that can help to survey the parameter space of
the full theory. For degenerate neutrino masses, the typical size of the
corrections is in the range of a one-loop threshold correction. In this
spirit, we applied this method of degeneracy-lifting to threshold
corrections as they typically occur in supersymmetric models including a
theory of neutrino masses. The \(\nul\)MSSM sets the stage of a very
powerful model. Although in its full generality, no statements about any
flavor pattern of the SUSY threshold corrections can ever be made, we
transferred the principle of degeneracy to the potentially arbitrary
parameters there and set the right-handed Majorana masses as well as the
neutrino Dirac masses to a degenerate pattern. In this very limited
setup, we looked for trilinear couplings \(\Mat{A}^\nul\) fulfilling the
requirements of the threshold corrections \(I_{ij}\) to end up at the
observed patterns of masses and mixings. The results are qualitatively
very stable under variation of the free SUSY parameters. In any case, we
need large neutrino \(A\) terms to get the structure of the threshold
corrections as for the generic discussion. Effectively, the combination
\(\Mat{A}^\nul/M_\upsh{SUSY}\) drives the corrections.

Shifting the generation of neutrino flavor from the mass matrices to the
soft SUSY breaking sector, especially \(\Mat{A}^\nul\), does not solve
the flavor puzzle neither reveals this procedure a deeper
understanding. The possibility of flavor mixing arising as a pure
quantum effect in the low energy effective theory, however,
opens another view on the flavor puzzle. Breaking Supersymmetry in a way
that does not respect flavor challenges also models for SUSY
breaking. Even if the UV extension of the SM is not supersymmetric, the
formulation of neutrino mixing via low-energy threshold corrections
implies a theory of flavor hidden in the yet veiled new physics.

\section*{Acknowledgments}
This work was supported by the DFG-funded research training group GRK
1694 (\emph{Elementarteilchenphysik bei h\"ochster Energie und
  h\"ochster Pr\"azision}). I acknowledge interesting and inspiring
discussions with S.~Pokorski which also triggered this work. Moreover, I
am very thankful for his useful and detailed comments on the first
version of the manuscript and his very thorough and patient reading of
the final version. I am also pleased about discussions with
U.~J.~Salda\~na Salazar and M.~Spinrath about this topic and flavor
mixing in general. I thank M.~Spinrath furthermore for reading and
commenting the manuscript.

\appendix
\input{app_sneumix}

\input{app_FeynmanRules}
\singlespacing
\bibliography{neutrino}
\end{document}

%% file: pack-def.tex
\usepackage{axodraw}
\usepackage[english]{babel}
\usepackage{cite}
\usepackage{a4wide}
\usepackage{amsmath}
\usepackage[charter,cal=cmcal,greekuppercase=italicized]{mathdesign}
\usepackage{dsfont}
\usepackage[T1]{fontenc}
\usepackage{slashed}
\usepackage{setspace}
\usepackage{graphicx}
\usepackage{xcolor}
\usepackage{soul}
\usepackage[final,          
            colorlinks,     
            linkcolor=purple, 
            citecolor=teal, 
            urlcolor=blue  
            ]{hyperref}

\makeatletter
\DeclareRobustCommand*{\bfseries}{%
  \not@math@alphabet\bfseries\mathbf
  \fontseries\bfdefault\selectfont
  \boldmath
}
\makeatother

\newcommand{\Ll}{\ensuremath{\mathrm{L}}}
\newcommand{\Rr}{\ensuremath{\mathrm{R}}}

\newcommand{\uq}{\ensuremath{\mathrm{u}}}
\newcommand{\el}{\ensuremath{\mathrm{e}}}
\newcommand{\nul}{\ensuremath{\nuup}}
\newcommand{\mul}{\ensuremath{\muup}}
\newcommand{\taul}{\ensuremath{\tauup}}

\newcommand{\upsh}[1]{\ensuremath{\mathrm{#1}}}
\newcommand{\im}{\ensuremath{\mathrm{i}\,}}

\newcommand{\diag}{\operatorname{diag}}
\newcommand{\SU}{\ensuremath{\mathrm{SU}}}

\newcommand{\SO}{\ensuremath{\mathrm{SO}}}
\newcommand{\CP}{\ensuremath{\mathrm{CP}}}
\newcommand{\eV}{\ensuremath{\mathrm{eV}}}
\newcommand{\GeV}{\ensuremath{\mathrm{GeV}}}
\newcommand{\TeV}{\ensuremath{\mathrm{TeV}}}
\newcommand{\MSbar}{\ensuremath{\overline{\mathrm{MS}}}}

\newcommand{\hc}{\ensuremath{\text{h.\,c.\ }}}
\newcommand{\tp}{\ensuremath{\text{T}}}
\renewcommand{\Re}{\ensuremath{\operatorname{Re}}}

\newcommand{\Mat}[1]{\ensuremath{\boldsymbol{#1}}}

\numberwithin{equation}{section}

%% file: app_sneumix.tex
\section{Sneutrino squared mass matrix and mixing matrix}\label{app:sneumix}

Extending the MSSM by right-handed neutrinos and giving them a Majorana
mass leads to a seesaw-like mechanism in the sneutrino sector. Similar
to the seesaw-extended Standard Model, where the neutrino spectrum gets
doubled, the sneutrino spectrum gets quadrupled. Why that? The MSSM
contains only three sneutrino states. Including right-handed fields, the
number of states get doubled, although half of them are singlets under
the SM gauge group. Moreover, due to Dirac and Majorana masses, the
physical spectrum gets even more enlarged. Effectively, we are left with
six light, more or less active states, and six heavy singlet-like
states. A priori, the sneutrino squared mass matrix is therefore a \(12
\times 12\) matrix, which can be perturbatively block-diagonalised
similar to the neutrino mass matrix. The complete procedure is described
in great detail by \cite{Dedes:2007ef}.

We choose the following basis: \(\tilde N = (\tilde\nu_\Ll,
\tilde\nu_\Ll^*, \tilde\nu_\Rr^*, \tilde\nu_\Rr)^\tp\) (such that
\(-\mathcal{L}^{\tilde\nul}_\upsh{mass} = N^\dag
(\mathcal{M}_{\mathbf{\tilde\nul}})^2 N\)) and classify chirality
conserving (\(LL, RR\)) and chirality changing blocks:
\begin{align*}
(\mathcal{M}_{\mathbf{\tilde\nu}})^2 =
\frac{1}{2}
        \begin{pmatrix}
                \mathcal{M}_{LL}^2 & \mathcal{M}_{LR}^2 \\
                \left(\mathcal{M}_{LR}^2\right)^\dag & \mathcal{M}_{RR}^2
        \end{pmatrix},
\end{align*}
with
\begin{subequations}
\begin{align}
  \mathcal{M}_{LL}^2 &= \begin{pmatrix}
      \mathcal{M}_{\tilde\ell}^2 + \frac{1}{2} M_Z^2 \cos2\beta
      \mathds{1}
      + \Mat{m}^\upsh{D}_\nul{\Mat{m}^\upsh{D}_\nul}^\dag & \Mat{0} \\
      \Mat{0} & (\mathcal{M}_{\tilde\ell}^2)^T
      + \frac{1}{2} M_Z^2 \cos2\beta \mathds{1}
      + {\Mat{m}^\upsh{D}_\nul}^* {\Mat{m}^\upsh{D}_\nul}^T
    \end{pmatrix},\\
  \mathcal{M}_{RL}^2 &= \begin{pmatrix}
      \Mat{m}^\upsh{D}_\nul\Mat{M}_\Rr &
      -\mu\cot\beta \Mat{m}^\upsh{D}_\nul + v_\uq \Mat{A}_\nul^* \\
      -\mu^* \cot\beta{\Mat{m}^\upsh{D}_\nul}^* + v_\uq \Mat{A}_\nul
      & {\Mat{m}^\upsh{D}_\nul}^* \Mat{M}_\Rr^*
    \end{pmatrix},\\
  \mathcal{M}_{RR}^2 &= \begin{pmatrix}
      (\mathcal{M}_{\tilde\nul}^2)^T +
      {\Mat{m}^\upsh{D}_\nul}^T{\Mat{m}^\upsh{D}_\nul}^*
      + \Mat{M}_\Rr^* \Mat{M}_\Rr & 2(\Mat{B}^2)^* \\
      2 \Mat{B}^2 & \mathcal{M}_{\tilde\nul}^2
      + {\Mat{m}^\upsh{D}_\nul}^\dag\Mat{m}^\upsh{D}_\nul
      + \Mat{M}_\Rr\Mat{M}^*_\Rr
    \end{pmatrix},
\end{align}
\end{subequations}
where bold face symbols as well as the soft slepton masses
\(\mathcal{M}^2_{\tilde \ell, \tilde\nul}\) denote \(3\times 3\)
matrices in flavour space and the singlet mass is symmetric:
\(\Mat{M}_\Rr = \Mat{M}_\Rr^T\). And \(\Mat{m}^\upsh{D}_\nul\)
is the Dirac neutrino mass matrix defined as
\(\Mat{m}^\upsh{D}_\nul=\frac{1}{\sqrt{2}} v_\uq \Mat{Y}_\nul\).

%% file: app_FeynmanRules.tex
\section{Feynman rules for the type I seesaw-extended MSSM}\label{app:FR}
The relevant vertices for the lepton flavor changing self energies are triple vertices for the lepton-slepton-gaugino and -higgsino interactions:
\begin{subequations}\label{eq:FRnuMSSM}
\begin{align}
  i\,\Gamma_{\nu_f}^{\tilde\nu_s \tilde\chi^0_k} = \, &
  -\frac{i}{\sqrt{2}} \bigg\lbrace \left[ (g_2 Z_{2k}^\upsh{N} - g_1
      Z_{1k}^\upsh{N}) \mathcal{Z}^{\tilde\nul *}_{is}
      (\Mat{U}_\upsh{PMNS})_{if} \right]
    P_\Ll \\ & \qquad\qquad + \left[(g_2 Z_{2k}^{\upsh{N}*} - g_1
      Z_{1k}^{\upsh{N}*}) \mathcal{Z}^{\tilde\nul}_{is}
      (\Mat{U}^*_\upsh{PMNS})_{if} \right] P_\Rr\bigg\rbrace ,\nonumber \\
  i\,\Gamma_{e_i}^{\tilde e_s \tilde\chi^0_k} =\, & \frac{i}{\sqrt{2}}
  \bigg\lbrace\left((g_2 Z_{2k}^\upsh{N} + g_1 Z_{1k}^\upsh{N})
      W^{\tilde \el}_{is} - y_{ij}^\el Z_{3k}^\upsh{N} W^{\tilde
        \el}_{j+3,s} \right) P_\Ll \\
    & \qquad\qquad- \left(\sqrt{2} g_1 Z_{1k}^\upsh{N} W^{\tilde
        \el}_{i+3,s} + y_{ji}^{\el*} Z_{3k}^\upsh{N} W^{\tilde
        \el}_{js}\right) P_\Rr \bigg\rbrace,\nonumber \\
  i\,\Gamma_{\nu_f}^{\tilde e_s \tilde\chi^+_k} =\, & -i \left[g_2
    Z_{1k}^- W^{\tilde \el*}_{is} - y_{ij}^{\el*} Z_{2k}^- W^{\tilde
      \el*}_{j+3,s} \right] (\Mat{U}_\upsh{PMNS})_{if} P_\Ll
  ,\\
  i\,\Gamma_{e_i}^{\tilde \nu_s \left(\tilde\chi^+_k\right)^c} =\, &
  -i \left[g_2 Z_{1k}^+ \mathcal{Z}^{\tilde \nul *}_{i,s} P_\Ll -
    y_{ij}^{\el} Z_{2k}^{-*} \mathcal{Z}^{\tilde\nul}_{js} P_\Rr
  \right],
\end{align}
\end{subequations}
where summation over double indices is understood.

The vertices of eqs.~\eqref{eq:FRnuMSSM} are given for an incoming
standard model fermion, outgoing chargino or neutralino as well as
sfermion. They generically follow from a interaction Lagrangian like
\[
\mathcal{L}_\mathrm{int} = \bar f_i \, \Gamma^{\tilde f_s
  \tilde\chi_k}_{f_f} \; \tilde\chi_k \, \tilde f_f \, + \mathrm{\ h.\
  c.}
\]

Each vertex comes along with the corresponding chirality projector:
\[
\Gamma^{\tilde f_s \tilde\chi_k}_{f_f} = \Gamma^{\tilde f_s
  \tilde\chi_k}_{\Ll, f_f} P_\Ll + \Gamma^{\tilde f_s
  \tilde\chi_k}_{\Rr, f_f} P_\Rr
\]

The mixing matrices diagonalize the mass matrices in the following manner:
\begin{itemize}
\item Chargino mixing: \(\Mat{Z}^{-\tp} \mathcal{M}_\upsh{C} \Mat{Z}^+ =
  \left(\mathcal{M}_\upsh{C}\right)^\upsh{diag}\),
 \item Neutralino mixing: \(\Mat{Z}^{\upsh{N}\tp} \mathcal{M}_\upsh{N}
   \Mat{Z}^\upsh{N} = \left(\mathcal{M}_\upsh{N}\right)^\upsh{diag}\),
 \item Slepton mixing: \(\Mat{W}^{\tilde \el \dag} \mathcal{M}^2_{\tilde \el}
   \Mat{W}^{\tilde \el} = \left(\mathcal{M}^2_{\tilde
       \el}\right)^\upsh{diag}\),
 \item Sneutrino mixing: \[\mathcal{W}^{\tilde\nul \dag}
   \bar{\mathcal{M}}_{\tilde\nul}^2 \mathcal{W}^{\tilde\nul} =
   \mathcal{W}^{\tilde\nul \dag} \mathcal{P}^\dag
   \mathcal{M}_{\tilde\nul}^2 \mathcal{P} \mathcal{W}^{\tilde\nul} =
   \left(\mathcal{M}^2_{\tilde\nul}\right)^\upsh{diag},\] such that
   \(\mathcal{Z}^{\tilde\nul} = \mathcal{P}\mathcal{W}^{\tilde\nul}\)
   diagonalizes the original mass matrix \(\mathcal{M}^2_{\tilde\nul}\)
   and therefore:
   \begin{align*}
     \mathcal{Z}_{is}^{\tilde \nul} &= \frac{1}{\sqrt{2}}
   \left( \mathcal{W}_{is}^{\tilde \nul} + i \mathcal{W}_{i+3,s}^{\tilde
       \nul}\right) \quad\text{and} \\
   \mathcal{Z}_{i+3,s}^{\tilde \nul} &=
   \frac{1}{\sqrt{2}} \left(\mathcal{W}_{is}^{\tilde \nul} - i
     \mathcal{W}_{i+3,s}^{\tilde \nul}\right)
   \end{align*}
   appear in the vertices of eqs.~\eqref{eq:FRnuMSSM}.
 \item Neutrino mixing: The PMNS mixing matrix can be determined from
   the neutrino mass matrix \(\Mat{m}_\nul^\upsh{diag} = \Mat{U}_\upsh{PMNS}^*
   \Mat{m}_\nul \Mat{U}_\upsh{PMNS}^\dag\), where \(m_\nul\) is the effective light
   neutrino mass matrix and the charged lepton masses can be taken
   diagonal (otherwise there would be a contribution to the PMNS mixing
   similar to the CKM mixing from both up and down sector:
   \(\Mat{U}_\upsh{PMNS} = \Mat{V}_{\el,\Ll}^\dag \Mat{U}_{\nul,\Ll}\), where
   \(\Mat{V}_{\el,\Ll}\) rotates the left-handed electron fields).
\end{itemize}

\paragraph*{Loop functions}
Finally, we give our conventions for the arising \(\MSbar\) subtracted
loop functions for \(p^2=0\) and \(\mu_r\) the renormalization scale:
\begin{subequations}
\begin{align}
  B_0(m_1, m_2) &= - \ln\frac{m_1 m_2}{\mu_r^2} + 1 -\frac{m_1^2 + m_2^2}{m_1^2 - m_2^2} \ln\frac{m_1}{m_2}, \\
        B_1(m_1, m_2) &= \frac{1}{2} \ln \frac{m_1 m_2}{\mu_r^2} - \frac{3}{4} - \frac{m_2^2}{2(m_1^2 - m_2^2)} + \left(\frac{m_1^4}{(m_1^2 - m_2^2)^2} - \frac{1}{2}\right) \frac{m_1}{m_2}.
\end{align}
\end{subequations}

%% file: degen_neutrinos.bbl
\providecommand{\href}[2]{#2}\begingroup\raggedright\begin{thebibliography}{10}

\bibitem{Osipowicz:2001sq}
{\bfseries KATRIN} Collaboration, A.~Osipowicz {\em et~al.}, ``{KATRIN: A Next
  generation tritium beta decay experiment with sub-eV sensitivity for the
  electron neutrino mass. Letter of intent},''
\href{http://arxiv.org/abs/hep-ex/0109033}{{\ttfamily arXiv:hep-ex/0109033
  [hep-ex]}}.

\bibitem{Abazajian:2011dt}
K.~Abazajian, E.~Calabrese, A.~Cooray, F.~De~Bernardis, S.~Dodelson, {\em
  et~al.}, ``{Cosmological and Astrophysical Neutrino Mass Measurements},''
  \href{http://dx.doi.org/10.1016/j.astropartphys.2011.07.002}{{\em
  Astropart.Phys.} {\bfseries 35} (2011) 177--184},
\href{http://arxiv.org/abs/1103.5083}{{\ttfamily arXiv:1103.5083
  [astro-ph.CO]}}.

\bibitem{Abazajian:2013oma}
K.~Abazajian {\em et~al.}, ``{Neutrino Physics from the Cosmic Microwave
  Background and Large Scale Structure},''
  \href{http://dx.doi.org/10.1016/j.astropartphys.2014.05.014}{{\em
  Astropart.Phys.} {\bfseries 63} (2014) 66--80},
\href{http://arxiv.org/abs/1309.5383}{{\ttfamily arXiv:1309.5383
  [astro-ph.CO]}}.

\bibitem{Ade:2013zuv}
{\bfseries \textsc{Planck}} Collaboration, P.~Ade {\em et~al.}, ``{Planck 2013
  results. XVI. Cosmological parameters},''
  \href{http://dx.doi.org/10.1051/0004-6361/201321591}{{\em Astron.Astrophys.}
  (2014) },
\href{http://arxiv.org/abs/1303.5076}{{\ttfamily arXiv:1303.5076
  [astro-ph.CO]}}.

\bibitem{Battye:2013xqa}
R.~A. Battye and A.~Moss, ``{Evidence for Massive Neutrinos from Cosmic
  Microwave Background and Lensing Observations},''
  \href{http://dx.doi.org/10.1103/PhysRevLett.112.051303}{{\em Phys.Rev.Lett.}
  {\bfseries 112} no.~5, (2014) 051303},
\href{http://arxiv.org/abs/1308.5870}{{\ttfamily arXiv:1308.5870
  [astro-ph.CO]}}.

\bibitem{Gonzalez-Garcia:2014bfa}
M.~Gonzalez-Garcia, M.~Maltoni, and T.~Schwetz, ``{Updated fit to three
  neutrino mixing: status of leptonic CP violation},''
\href{http://arxiv.org/abs/1409.5439}{{\ttfamily arXiv:1409.5439 [hep-ph]}}.

\bibitem{Ellis:1999my}
J.~R. Ellis and S.~Lola, ``{Can neutrinos be degenerate in mass?},''
  \href{http://dx.doi.org/10.1016/S0370-2693(99)00545-6}{{\em Phys.Lett.}
  {\bfseries B458} (1999) 310--321},
\href{http://arxiv.org/abs/hep-ph/9904279}{{\ttfamily arXiv:hep-ph/9904279
  [hep-ph]}}.

\bibitem{Casas:1999tp}
J.~Casas, J.~Espinosa, A.~Ibarra, and I.~Navarro, ``{Naturalness of nearly
  degenerate neutrinos},''
  \href{http://dx.doi.org/10.1016/S0550-3213(99)00383-1}{{\em Nucl.Phys.}
  {\bfseries B556} (1999) 3--22},
\href{http://arxiv.org/abs/hep-ph/9904395}{{\ttfamily arXiv:hep-ph/9904395
  [hep-ph]}}.

\bibitem{Casas:1999ac}
J.~Casas, J.~Espinosa, A.~Ibarra, and I.~Navarro, ``{Nearly degenerate
  neutrinos, supersymmetry and radiative corrections},''
  \href{http://dx.doi.org/10.1016/S0550-3213(99)00605-7}{{\em Nucl.Phys.}
  {\bfseries B569} (2000) 82--106},
\href{http://arxiv.org/abs/hep-ph/9905381}{{\ttfamily arXiv:hep-ph/9905381
  [hep-ph]}}.

\bibitem{Haba:1999xz}
N.~Haba, Y.~Matsui, N.~Okamura, and M.~Sugiura, ``{The Effect of Majorana phase
  in degenerate neutrinos},'' \href{http://dx.doi.org/10.1143/PTP.103.145}{{\em
  Prog.Theor.Phys.} {\bfseries 103} (2000) 145--150},
\href{http://arxiv.org/abs/hep-ph/9908429}{{\ttfamily arXiv:hep-ph/9908429
  [hep-ph]}}.

\bibitem{Chun:1999vb}
E.~J. Chun and S.~Pokorski, ``{Slepton flavor mixing and neutrino masses},''
  \href{http://dx.doi.org/10.1103/PhysRevD.62.053001}{{\em Phys.Rev.}
  {\bfseries D62} (2000) 053001},
\href{http://arxiv.org/abs/hep-ph/9912210}{{\ttfamily arXiv:hep-ph/9912210
  [hep-ph]}}.

\bibitem{Balaji:2000gd}
K.~Balaji, A.~S. Dighe, R.~Mohapatra, and M.~Parida, ``{Generation of large
  flavor mixing from radiative corrections},''
  \href{http://dx.doi.org/10.1103/PhysRevLett.84.5034}{{\em Phys.Rev.Lett.}
  {\bfseries 84} (2000) 5034--5037},
\href{http://arxiv.org/abs/hep-ph/0001310}{{\ttfamily arXiv:hep-ph/0001310
  [hep-ph]}}.

\bibitem{Chankowski:2000fp}
P.~H. Chankowski, A.~Ioannisian, S.~Pokorski, and J.~Valle, ``{Neutrino
  unification},'' \href{http://dx.doi.org/10.1103/PhysRevLett.86.3488}{{\em
  Phys.Rev.Lett.} {\bfseries 86} (2001) 3488--3491},
\href{http://arxiv.org/abs/hep-ph/0011150}{{\ttfamily arXiv:hep-ph/0011150
  [hep-ph]}}.

\bibitem{Chun:2001kh}
E.~J. Chun, ``{Lepton flavor violation and radiative neutrino masses},''
  \href{http://dx.doi.org/10.1016/S0370-2693(01)00387-2}{{\em Phys.Lett.}
  {\bfseries B505} (2001) 155--160},
\href{http://arxiv.org/abs/hep-ph/0101170}{{\ttfamily arXiv:hep-ph/0101170
  [hep-ph]}}.

\bibitem{Chankowski:2001mx}
P.~H. Chankowski and S.~Pokorski, ``{Quantum corrections to neutrino masses and
  mixing angles},'' \href{http://dx.doi.org/10.1142/S0217751X02006109}{{\em
  Int.J.Mod.Phys.} {\bfseries A17} (2002) 575--614},
\href{http://arxiv.org/abs/hep-ph/0110249}{{\ttfamily arXiv:hep-ph/0110249
  [hep-ph]}}.

\bibitem{Mohapatra:2003tw}
R.~Mohapatra, M.~Parida, and G.~Rajasekaran, ``{High scale mixing unification
  and large neutrino mixing angles},''
  \href{http://dx.doi.org/10.1103/PhysRevD.69.053007}{{\em Phys.Rev.}
  {\bfseries D69} (2004) 053007},
\href{http://arxiv.org/abs/hep-ph/0301234}{{\ttfamily arXiv:hep-ph/0301234
  [hep-ph]}}.

\bibitem{Brahmachari:2003dj}
B.~Brahmachari and E.~J. Chun, ``{Supersymmetric threshold corrections to
  \(\Delta m^2_\odot\)},''
  \href{http://dx.doi.org/10.1016/j.physletb.2004.03.004}{{\em Phys.Lett.}
  {\bfseries B596} (2004) 184--190},
\href{http://arxiv.org/abs/hep-ph/0312030}{{\ttfamily arXiv:hep-ph/0312030
  [hep-ph]}}.

\bibitem{Mohapatra:2005gs}
R.~Mohapatra, M.~Parida, and G.~Rajasekaran, ``{Threshold effects on
  quasi-degenerate neutrinos with high-scale mixing unification},''
  \href{http://dx.doi.org/10.1103/PhysRevD.71.057301}{{\em Phys.Rev.}
  {\bfseries D71} (2005) 057301},
\href{http://arxiv.org/abs/hep-ph/0501275}{{\ttfamily arXiv:hep-ph/0501275
  [hep-ph]}}.

\bibitem{Haba:2012ar}
N.~Haba and R.~Takahashi, ``{Grand Unification of Flavor Mixings},''
  \href{http://dx.doi.org/10.1209/0295-5075/100/31001}{{\em Europhys.Lett.}
  {\bfseries 100} (2012) 31001},
\href{http://arxiv.org/abs/1206.2793}{{\ttfamily arXiv:1206.2793 [hep-ph]}}.

\bibitem{Minkowski:1977sc}
P.~Minkowski, ``{\(\mu \to e \gamma\) at a Rate of One Out of 1-Billion Muon
  Decays?},''
\href{http://dx.doi.org/10.1016/0370-2693(77)90435-X}{{\em Phys.Lett.}
  {\bfseries B67} (1977) 421}.

\bibitem{Mohapatra:1979ia}
R.~N. Mohapatra and G.~Senjanovic, ``{Neutrino Mass and Spontaneous Parity
  Violation},'' \href{http://dx.doi.org/10.1103/PhysRevLett.44.912}{{\em
  Phys.Rev.Lett.} {\bfseries 44} (1980) 912}.

\bibitem{Yanagida:1980xy}
T.~Yanagida, ``{Horizontal Symmetry and Masses of Neutrinos},''
\href{http://dx.doi.org/10.1143/PTP.64.1103}{{\em Prog.Theor.Phys.} {\bfseries
  64} (1980) 1103}.

\bibitem{GellMann:1980vs}
M.~Gell-Mann, P.~Ramond, and R.~Slansky, ``{Complex Spinors and Unified
  Theories},'' {\em Conf.Proc.} {\bfseries C790927} (1979) 315--321,
\href{http://arxiv.org/abs/1306.4669}{{\ttfamily arXiv:1306.4669 [hep-th]}}.

\bibitem{Schechter:1980gr}
J.~Schechter and J.~Valle, ``{Neutrino Masses in SU(2) x U(1) Theories},''
\href{http://dx.doi.org/10.1103/PhysRevD.22.2227}{{\em Phys.Rev.} {\bfseries
  D22} (1980) 2227}.

\bibitem{Magg:1980ut}
M.~Magg and C.~Wetterich, ``{Neutrino Mass Problem and Gauge Hierarchy},''
\href{http://dx.doi.org/10.1016/0370-2693(80)90825-4}{{\em Phys.Lett.}
  {\bfseries B94} (1980) 61}.

\bibitem{Schechter:1981cv}
J.~Schechter and J.~Valle, ``{Neutrino Decay and Spontaneous Violation of
  Lepton Number},''
\href{http://dx.doi.org/10.1103/PhysRevD.25.774}{{\em Phys.Rev.} {\bfseries
  D25} (1982) 774}.

\bibitem{Weinberg:1979sa}
S.~Weinberg, ``{Baryon and Lepton Nonconserving Processes},''
\href{http://dx.doi.org/10.1103/PhysRevLett.43.1566}{{\em Phys.Rev.Lett.}
  {\bfseries 43} (1979) 1566--1570}.

\bibitem{Pontecorvo:1957qd}
B.~Pontecorvo, ``{Inverse beta processes and nonconservation of lepton
  charge},''
{\em Sov.Phys.JETP} {\bfseries 7} (1958) 172--173.

\bibitem{Maki:1962mu}
Z.~Maki, M.~Nakagawa, and S.~Sakata, ``{Remarks on the unified model of
  elementary particles},''
\href{http://dx.doi.org/10.1143/PTP.28.870}{{\em Prog.Theor.Phys.} {\bfseries
  28} (1962) 870--880}.

\bibitem{Chankowski:1993tx}
P.~H. Chankowski and Z.~Pluciennik, ``{Renormalization group equations for
  seesaw neutrino masses},''
  \href{http://dx.doi.org/10.1016/0370-2693(93)90330-K}{{\em Phys.Lett.}
  {\bfseries B316} (1993) 312--317},
\href{http://arxiv.org/abs/hep-ph/9306333}{{\ttfamily arXiv:hep-ph/9306333
  [hep-ph]}}.

\bibitem{Haba:1998fb}
N.~Haba, N.~Okamura, and M.~Sugiura, ``{The Renormalization group analysis of
  the large lepton flavor mixing and the neutrino mass},''
  \href{http://dx.doi.org/10.1143/PTP.103.367}{{\em Prog.Theor.Phys.}
  {\bfseries 103} (2000) 367--377},
\href{http://arxiv.org/abs/hep-ph/9810471}{{\ttfamily arXiv:hep-ph/9810471
  [hep-ph]}}.

\bibitem{Casas:1999tg}
J.~Casas, J.~Espinosa, A.~Ibarra, and I.~Navarro, ``{General RG equations for
  physical neutrino parameters and their phenomenological implications},''
  \href{http://dx.doi.org/10.1016/S0550-3213(99)00781-6}{{\em Nucl.Phys.}
  {\bfseries B573} (2000) 652--684},
\href{http://arxiv.org/abs/hep-ph/9910420}{{\ttfamily arXiv:hep-ph/9910420
  [hep-ph]}}.

\bibitem{Chankowski:1999xc}
P.~H. Chankowski, W.~Krolikowski, and S.~Pokorski, ``{Fixed points in the
  evolution of neutrino mixings},''
  \href{http://dx.doi.org/10.1016/S0370-2693(99)01465-3}{{\em Phys.Lett.}
  {\bfseries B473} (2000) 109--117},
\href{http://arxiv.org/abs/hep-ph/9910231}{{\ttfamily arXiv:hep-ph/9910231
  [hep-ph]}}.

\bibitem{Haba:2000tx}
N.~Haba, Y.~Matsui, and N.~Okamura, ``{The Effects of Majorana phases in three
  generation neutrinos},'' \href{http://dx.doi.org/10.1007/s100520000458}{{\em
  Eur.Phys.J.} {\bfseries C17} (2000) 513--520},
\href{http://arxiv.org/abs/hep-ph/0005075}{{\ttfamily arXiv:hep-ph/0005075
  [hep-ph]}}.

\bibitem{Wolfenstein:1981rk}
L.~Wolfenstein, ``{CP Properties of Majorana Neutrinos and Double beta
  Decay},''
\href{http://dx.doi.org/10.1016/0370-2693(81)91151-5}{{\em Phys.Lett.}
  {\bfseries B107} (1981) 77}.

\bibitem{Branco:1998bw}
G.~Branco, M.~Rebelo, and J.~Silva-Marcos, ``{Degenerate and quasidegenerate
  Majorana neutrinos},''
  \href{http://dx.doi.org/10.1103/PhysRevLett.82.683}{{\em Phys.Rev.Lett.}
  {\bfseries 82} (1999) 683--686},
\href{http://arxiv.org/abs/hep-ph/9810328}{{\ttfamily arXiv:hep-ph/9810328
  [hep-ph]}}.

\bibitem{Branco:2014zza}
G.~Branco, M.~Rebelo, J.~Silva-Marcos, and D.~Wegman, ``{Quasidegeneracy of
  Majorana Neutrinos and the Origin of Large Leptonic Mixing},''
\href{http://arxiv.org/abs/1405.5120}{{\ttfamily arXiv:1405.5120 [hep-ph]}}.

\bibitem{GonzalezGarcia:2012sz}
M.~Gonzalez-Garcia, M.~Maltoni, J.~Salvado, and T.~Schwetz, ``{Global fit to
  three neutrino mixing: critical look at present precision},''
  \href{http://dx.doi.org/10.1007/JHEP12(2012)123}{{\em JHEP} {\bfseries 1212}
  (2012) 123},
\href{http://arxiv.org/abs/1209.3023}{{\ttfamily arXiv:1209.3023 [hep-ph]}}.

\bibitem{Chankowski:2001hx}
P.~H. Chankowski and P.~Wasowicz, ``{Low-energy threshold corrections to
  neutrino masses and mixing angles},''
  \href{http://dx.doi.org/10.1007/s100520100867}{{\em Eur.Phys.J.} {\bfseries
  C23} (2002) 249--258},
\href{http://arxiv.org/abs/hep-ph/0110237}{{\ttfamily arXiv:hep-ph/0110237
  [hep-ph]}}.

\bibitem{Babu:2002dz}
K.~Babu, E.~Ma, and J.~Valle, ``{Underlying A(4) symmetry for the neutrino mass
  matrix and the quark mixing matrix},''
  \href{http://dx.doi.org/10.1016/S0370-2693(02)03153-2}{{\em Phys.Lett.}
  {\bfseries B552} (2003) 207--213},
\href{http://arxiv.org/abs/hep-ph/0206292}{{\ttfamily arXiv:hep-ph/0206292
  [hep-ph]}}.

\bibitem{Morisi:2013qna}
S.~Morisi, D.~Forero, J.~Romão, and J.~Valle, ``{Neutrino mixing with revamped
  $A_4$ flavor symmetry},''
  \href{http://dx.doi.org/10.1103/PhysRevD.88.016003}{{\em Phys.Rev.}
  {\bfseries D88} no.~1, (2013) 016003},
\href{http://arxiv.org/abs/1305.6774}{{\ttfamily arXiv:1305.6774 [hep-ph]}}.

\bibitem{Hollik:2014jda}
W.~G. Hollik and U.~J.~S. Salazar, ``{The double mass hierarchy pattern:
  simultaneously understanding quark and lepton mixing},''
\href{http://arxiv.org/abs/1411.3549}{{\ttfamily arXiv:1411.3549 [hep-ph]}}.

\bibitem{Farzan:2004cm}
Y.~Farzan, ``{Effects of the neutrino B-term on the Higgs mass parameters and
  electroweak symmetry breaking},''
  \href{http://dx.doi.org/10.1088/1126-6708/2005/02/025}{{\em JHEP} {\bfseries
  0502} (2005) 025},
\href{http://arxiv.org/abs/hep-ph/0411358}{{\ttfamily arXiv:hep-ph/0411358
  [hep-ph]}}.

\bibitem{Dedes:2007ef}
A.~Dedes, H.~E. Haber, and J.~Rosiek, ``{Seesaw mechanism in the sneutrino
  sector and its consequences},''
  \href{http://dx.doi.org/10.1088/1126-6708/2007/11/059}{{\em JHEP} {\bfseries
  0711} (2007) 059},
\href{http://arxiv.org/abs/0707.3718}{{\ttfamily arXiv:0707.3718 [hep-ph]}}.

\bibitem{Heinemeyer:2014hka}
S.~Heinemeyer, J.~Hernandez-Garcia, M.~Herrero, X.~Marcano, and
  A.~Rodriguez-Sanchez, ``{Radiative corrections to $M_h$ from three
  generations of Majorana neutrinos and sneutrinos},''
\href{http://arxiv.org/abs/1407.1083}{{\ttfamily arXiv:1407.1083 [hep-ph]}}.

\bibitem{Casas:2001sr}
J.~Casas and A.~Ibarra, ``{Oscillating neutrinos and \(\mu \to e, \gamma\)},''
  \href{http://dx.doi.org/10.1016/S0550-3213(01)00475-8}{{\em Nucl.Phys.}
  {\bfseries B618} (2001) 171--204},
\href{http://arxiv.org/abs/hep-ph/0103065}{{\ttfamily arXiv:hep-ph/0103065
  [hep-ph]}}.

\bibitem{Davidson:2008bu}
S.~Davidson, E.~Nardi, and Y.~Nir, ``{Leptogenesis},''
  \href{http://dx.doi.org/10.1016/j.physrep.2008.06.002}{{\em Phys.Rept.}
  {\bfseries 466} (2008) 105--177},
\href{http://arxiv.org/abs/0802.2962}{{\ttfamily arXiv:0802.2962 [hep-ph]}}.

\end{thebibliography}\endgroup
